\documentclass[aps,prd,longbibliography,reprint,twocolumn,amsmath,amssymb,amsfonts,showpacs,footnote]{revtex4-1}

\usepackage{ifpdf}
\usepackage{dcolumn}
\usepackage{enumitem}
\usepackage{bm}
\usepackage{here}
\usepackage{float}

\floatstyle{plaintop}
\restylefloat{table}
\ifpdf

      \usepackage[pdftex]{graphicx}  
      \usepackage[pdftex]{epsfig}

      \usepackage[pdftex]{hyperref}
\hypersetup
	{
		colorlinks,%
		citecolor=blue,%
		linkcolor=red,%
		urlcolor=blue,%
	}

\else

      \usepackage[dvips]{graphicx}  
      \usepackage[dvips]{epsfig}

      \usepackage[dvipdfm]{hyperref}
      \hypersetup
	{
		colorlinks,%
		citecolor=blue,%
		linkcolor=red,%
		urlcolor=blue,%
	}

\fi

\begin{document}

\title{An alternative description of gravitational radiation from black holes \\
based on the Regge poles of the ${\cal S}$-matrix and the associated residues}

\author{Antoine Folacci}
\email{folacci@univ-corse.fr}
\affiliation{Equipe Physique
Th\'eorique, \\ SPE, UMR 6134 du CNRS
et de l'Universit\'e de Corse,\\
Universit\'e de Corse, Facult\'e des Sciences, BP 52, F-20250 Corte,
France}

\author{Mohamed \surname{Ould~El~Hadj}}
\email{med.ouldelhadj@gmail.com}
\affiliation{Equipe Physique
Th\'eorique, \\ SPE, UMR 6134 du CNRS
et de l'Universit\'e de Corse,\\
Universit\'e de Corse, Facult\'e des Sciences, BP 52, F-20250 Corte,
France}

\date{\today}

\begin{abstract}

We advocate for an alternative description of gravitational radiation from black holes based on complex angular momentum techniques (analytic continuation of partial wave expansions, duality of the ${\cal S}$-matrix and effective resummations involving its Regge poles and the associated residues, Regge trajectories, semiclassical interpretations, etc.). Such techniques, which proved to be very helpful in various areas of physics to describe and analyze resonant scattering, were only marginally used in the context of black hole physics. Here, by considering the multipolar waveform generated by a massive particle falling radially from infinity into a Schwarzschild black hole, we show that they could play a fundamental role in gravitational-wave physics. More precisely, from the multipole expansion defining the Weyl scalar $\Psi_4$, we extract the Fourier transform of a sum over Regge poles and their residues which can be evaluated numerically from the associated Regge trajectories. This Regge pole approximation permits us to reconstruct, for an arbitrary direction of observation, a large part of the multipolar waveform $\Psi_4$. In particular, it can reproduce with very good agreement the quasinormal ringdown as well as with rather good agreement the tail of the signal. This is achieved even if we take into account only one Regge pole and, if a large number of modes are excited, the result can be improved by considering additional poles. Moreover, while quasinormal-mode contributions do not provide physically relevant results at ``early times'' due to their exponentially divergent behavior as time decreases, it is not necessary to determine from physical considerations a starting time for the Regge ringdown.

\end{abstract}

\maketitle

\tableofcontents

\section{Introduction}

In the mid-1970s, Chandrasekhar observed that the study of black hole (BH) perturbations can be reduced to a problem of resonant scattering (cf.~Ref.~\cite{Chandrasekhar:1985kt} and references therein). This point of view puts at the heart of BH perturbation theory the ${\cal S}$-matrix concept. As a consequence, it is an invitation to use systematically, in the context of BH physics, the various tools developed in the framework of resonant scattering theory and, in particular, to fully exploit the dual structure of the ${\cal S}$-matrix; indeed, such a matrix is a double-entry mathematical object that is a function of both the angular momentum $\ell \in \mathbb{N}$ and the frequency $\omega \in \mathbb{R}$ and that can be analytically extended (i) for $\ell \in \mathbb{N}$, into the complex $\omega$ plane and (ii) for $\omega \in \mathbb{R}$, into the complex $\ell$ plane [the so-called complex angular momentum (CAM) plane]. It is important to note that this duality permits us to shed light, from two different points of view, on a resonant phenomenon and to juggle with its two alternative descriptions.

It is well known that the analytic structure of the ${\cal S}$-matrix in the complex $\omega$ plane permits us to physically interpret the response of a BH to an external excitation. This was developed in a seminal paper by Leaver \cite{Leaver:1986gd}. In particular, (i) the poles of the ${\cal S}$-matrix and the associated residues are, respectively, the complex frequencies and the excitation factors of the quasinormal modes (QNMs) which are involved in the description of the BH ringdown, that part of the signal that dominates the BH response at intermediate timescales while (ii) a branch-cut integral allows us to describe the tail of the signal, i.e., the BH response at very late times. Such a point of view is now widely considered in the literature. On the other hand, there is very little work based on the analytic structure of the ${\cal S}$-matrix in the complex $\ell$ plane. This is really surprising. Indeed, in all the other areas of physics involving resonant scattering theory (see, e.g., Refs.~\cite{deAlfaro:1965zz,Newton:1982qc} for quantum mechanics, Refs.~\cite{Watson18,Sommerfeld49,Newton:1982qc,Nus92,Grandy2000} for electromagnetism and optics, Refs.~\cite{Uberall1992,AkiRichards2002} for acoustics and seismology, and Refs.~\cite{Gribov69,Collins77,BaronePredazzi2002,DonnachieETAL2005} for high energy physics), it is common to analyze physical phenomenons by using CAM techniques and by considering the poles of the ${\cal S}$-matrix in the CAM plane (the so-called Regge poles). Such techniques are very helpful because they permit us to extract the physical information encoded into partial wave expansions by providing (i) powerful tools of resummation of these expansions and (ii) ``semiclassical'' descriptions of resonance phenomenons.

Even if, as we have just noted, the CAM approach has been neglected in the context of BH physics in favor of descriptions based on the analytic structure of the ${\cal S}$-matrix in the complex $\omega$ plane, there exists, however, a few works using CAM techniques. Among these works, we could quote the following ones because they are more or less relevant to the present article:

\begin{enumerate}[label=(\arabic*)]

   \item  Chandrasekar and Ferrari have used the Regge pole theory to determine the flow of gravitational energy through a relativistic star \cite{Chandrasekhar:1992ey}.

   \item  Andersson and Thylwe have provided a CAM description of the scattering of monochromatic scalar waves by a Schwarzschild BH and used it to interpret the BH glory \cite{Andersson:1994rk,Andersson:1994rm}. They have, in particular, considered ``surface waves'' propagating close to the unstable circular photon (graviton) orbit at $r=3M$ (i.e., near the so-called photon sphere) and associated them with the Regge poles of the ${\cal S}$-matrix.

\item  For the Schwarzschild BH, we have established that the complex frequencies of the weakly damped QNMs are Breit-Wigner-type resonances generated by the surface waves previously mentioned and we have been able to construct semiclassically the spectrum of the QNM complex frequencies from the Regge trajectories, i.e., from the curves traced out in the CAM plane by the Regge poles as a function of the frequency \cite{Decanini:2002ha,Decanini:2009mu}. In this way, we have established on a ``rigorous'' basis the physically intuitive interpretation of the Schwarzschild BH QNMs suggested, as early as 1972, by Goebel \cite{Goebel}. These results have been extended to other BHs or massive fields in Refs.~\cite{Decanini:2009dn,Dolan:2009nk,Decanini:2010fz,Decanini:2011eh}.

\item From the Regge trajectories and the residues of the greybody factors, we have described analytically the high-energy absorption cross section for a wide class of BHs endowed with a photon sphere and explained its oscillations in terms of the geometrical characteristics (orbital period and Lyapunov exponent) of the null unstable geodesics lying on the photon sphere \cite{Decanini:2011xi,Decanini:2011xw,Decanini:2011eh}.

\end{enumerate}

In this article, we show that the CAM machinery provides an interesting alternative framework for describing gravitational radiation from perturbed BHs. For that, we consider the classical problem of a Schwarzschild BH perturbed by a massive particle falling radially from infinity with an arbitrary initial kinetic energy. Since the pioneering works, in the 1970s, concerning this problem \cite{Zerilli:1971wd,Davis:1971gg,Davis:1972ud,Tiomno:1972dq,Ruffini:1972uh,Ruffini:1973ky,Ferrari:1981dh}, it has been regularly revisited (cf., e.g., Refs.~\cite{Leaver:1986gd,Lousto:1996sx,Lousto:1997wf,
Martel:2001yf,Cardoso:2002ay,Cardoso:2002yj,Cardoso:2002jr,Cardoso:2003cn,Mitsou:2010jv,Berti:2010ce,Berti:2010gx,
Sperhake:2011ik,Degollado:2014dfa,Moreno:2016urq,Aranha:2016sgg,Oliveira:2018pnx}) due to its canonical importance (it permits us to discuss theoretical concepts or to test numerical tools) and also because it provides a simplified model for head-on collisions of BHs. Here, we describe the gravitational radiation from the Weyl scalar $\Psi_4$ and extract from its multipole expansion the Fourier transform of a sum over Regge poles and their residues which can be evaluated numerically from the associated Regge trajectories. This Regge pole approximation of the Weyl scalar $\Psi_4$ permits us to construct, for an arbitrary direction of observation, a large part of the multipolar waveform. In particular, it can reproduce with very good (sometimes impressive) agreement the quasinormal ringdown as well as with rather good agreement the tail of the signal. This is achieved even if we take into account only one Regge pole and these agreements can be improved by considering additional Regge poles. In fact, using Regge poles to describe the ringdown is equivalent to extracting the information encoded into the full quasinormal-mode spectrum by resumming over a large number of quasinormal frequencies and the associated quasinormal excitation factors. Moreover, it is interesting to note that, while QNM contributions do not provide physically relevant results at early times due to their exponentially divergent behavior as time decreases, it is not necessary to determine from physical considerations a starting time for the Regge ringdown.

Our paper is organized as follows. In Sec.~\ref{SecII}, we first construct the Weyl scalar $\Psi_4$ describing the outgoing radiation at infinity which is generated by the massive particle falling radially into a Schwarzschild BH. This is achieved by solving, in the frequency domain and from standard Green's function techniques, the Zerilli-Moncrief equation \cite{Zerilli:1971wd,Moncrief:1974am} for arbitrary $(\ell,m)$ partial modes. Here, we consider both the case of a particle starting at rest from infinity and of a particle projected with a finite kinetic energy at infinity. We also extract from the multipole expansion of $\Psi_4$ the quasinormal ringdown of the BH. In Sec.~\ref{SecIII}, by means of the Poisson summation formula \cite{MorseFeshbach1953}, and also by means of the Sommerfeld-Watson transform \cite{Watson18,Sommerfeld49,Newton:1982qc} and Cauchy's theorem, we provide two different CAM representations of the multipolar waveform $\Psi_4$ from which we extract the Fourier transform of a sum over Regge poles and their residues. These two Regge pole approximations of the Weyl scalar $\Psi_4$ can be evaluated numerically if we have at our disposal the Regge trajectories. Here, by Regge trajectories, we intend the curves traced out in the CAM plane by the Regge poles, as well as by the associated residues, as a function of the frequency $\omega$. We obtain numerically these Regge trajectories in this section. In Sec.~\ref{SecIV}, we compare numerically the multipolar waveform $\Psi_4$ constructed by summing over a large number of partial modes (this is particularly necessary for a radially infalling relativistic particle) as well as the associated quasinormal ringdown with the Regge pole approximations obtained in Sec.~\ref{SecIII}. This allows us to clearly highlight the benefits of working with the Regge pole approximations of $\Psi_4$. In the Conclusion, we summarize our main results, briefly consider possible extensions of our work, and return to the philosophy underlying the CAM approach of BH physics.

Throughout this article, we adopt units such that $G = c = 1$ and we use the geometrical conventions of Ref.~\cite{Misner:1974qy}. We, furthermore, consider that the exterior of the Schwarzschild BH is defined by the line element $ds^2= -f(r) dt^2+ f(r)^{-1}dr^2+ r^2 d\theta^2 + r^2 \sin^2\theta d\varphi^2$ where $f(r)=1-2M/r$ and $M$ is the mass of the BH while $t \in ]-\infty, +\infty[$, $r \in ]2M,+\infty[$, $\theta \in [0,\pi]$ and $\varphi \in [0,2\pi]$ are the usual Schwarzschild coordinates.

\section{Weyl scalar $\Psi_4$ and associated quasinormal ringdown}
\label{SecII}

In this section, we shall construct the Weyl scalar $\Psi_4$ describing the outgoing radiation at infinity due to a point particle of mass $m_0$ falling radially from infinity into a Schwarzschild BH. Moreover, we shall extract from the multipole expansion of $\Psi_4$ the associated ringdown waveform.

\subsection{Multipole expansion of the Weyl scalar $\Psi_4$}
\label{SecIIa}

We assume an extreme mass ratio for the system particle-BH (i.e., that $m_0 \ll M$), such a hypothesis permitting us to describe the gravitational radiation in the framework of BH perturbations (see, for pioneering works on this topic, the articles by Regge and Wheeler \cite{Regge:1957td}, Zerilli \cite{Zerilli:1970se,Zerilli:1971wd} and Vishveshwara \cite{Vishveshwara:1970cc} as well as the article by Martel and Poisson \cite{Martel:2005ir} and the review by Nagar and Rezzolla \cite{Nagar:2005ea} for a modern point of view based on gauge invariance \cite{Moncrief:1974am,Cunningham:1978zfa,Cunningham:1979px}. Moreover, we recall that, in the context of BH perturbation theory, it is possible to describe the radiation contained in the emitted gravitational waves in the framework of the Newman-Penrose formalism by means of the Weyl scalar $\Psi_4$ \cite{Newman:1961qr,Chandrasekhar:1985kt}. We refer more particularly to Chap.~8 of Ref.~\cite{Alcubierre:1138167} for a short review of the Newman-Penrose formalism and its use in connection with the perturbations of the Schwarzschild BH and we note that, in this context, the Weyl scalar $\Psi_4$ can be expressed for $r \to +\infty$ as
\begin{equation}\label{Psi4_def}
\Psi_4 = \frac{1}{2} \frac{\partial^2}{\partial t^2} \left( h_{+} - i  h_{\times} \right).
\end{equation}
Here, $\Psi_4$ has been defined with respect to the null basis $(l,n,m,m^\ast)$ which is normalized such that the only nonvanishing scalar products involving the vectors of the tetrad are $l^\mu n_\mu = -1$ and $m^\mu m^\ast_\mu=1$ and which is given by (it should be noted that our conventions slightly differ from those of Ref.~\cite{Alcubierre:1138167})
\begin{subequations}
\begin{eqnarray} \label{NullTetrad}
& & l^\mu=\left(\frac{1}{f(r)},1,0,0 \right), \\
& & n^\mu=\frac{1}{2}\left(1,-\frac{1}{f(r)},0,0\right), \\
& & m^\mu=\frac{1}{\sqrt{2} r}\left(0,0,1, \frac{i}{\sin \theta}\right), \\
& & {m^\ast}^\mu=\frac{1}{\sqrt{2} r}\left(0,0,1, \frac{-i}{\sin \theta}\right).
\end{eqnarray}
\end{subequations}
In the right-hand side (r.h.s.)~of Eq.~(\ref{Psi4_def}), $h_{+}$ and $h_{\times}$ denote, in the transverse traceless gauge, the two circularly polarized components of the emitted gravitational wave \cite{Misner:1974qy}.

The timelike geodesic followed by the particle falling radially into the Schwarzschild BH is defined by the coordinates $t_{p}(\tau)$, $r_{p}(\tau)$, $\theta_{p}(\tau)$ and $\varphi_{p}(\tau)$ where $\tau$ is the proper time of the particle. Without loss of generality, we can consider that the particle moves in the BH equatorial plane along the positive $x$ axis and in the negative direction, i.e., we assume that $\theta_{p}(\tau)=\pi/2$, $\varphi_{p}(\tau)=0$ and $dr_{p}(\tau)/{d\tau} <0$. The functions $t_{p}(\tau)$, $r_{p}(\tau)$ as well as the function $t_{p}(r)$ can be then obtained from the geodesic equations (see, e.g., Ref.~\cite{Chandrasekhar:1985kt})
\begin{subequations}
\label{geodesic_equations}
\begin{equation}
f(r_{p})\frac{dt_{p}}{d\tau}=\frac{E}{m_0}, \label{geodesic_1}
\end{equation}
\noindent and
\begin{equation}
\label{geodesic_3}
\left(\frac{dr_{p}}{d\tau}\right)^{2} -\frac{2M}{r_{p}}
=\left(\frac{E}{m_0}\right)^{2}-1.
\end{equation}
\end{subequations}
Here $E$ is the energy of the particle. It is a constant of motion which can be related to the velocity $v_\infty$ of the particle at infinity and to the associated Lorentz factor $\gamma$ by
\begin{equation}
\label{tildeEand gamma}
\frac{E}{m_0}=\frac{1}{\sqrt{1-(v_\infty)^2}}=\gamma.
\end{equation}

We also recall that the radially infalling particle only excites the even (polar) modes of the Schwarzschild BH and that, in the transverse traceless gauge, the two circularly polarized components $(h_{+},h_{\times})$ of the emitted gravitational wave can be expanded on the (scalar) spherical harmonics $Y^{\ell m}(\theta,\varphi)$ in the form \cite{Nagar:2005ea}
\begin{subequations} \label{hp_hc}
\begin{eqnarray}
& & h_{+}= \frac{1}{r} \sum_{\ell=2}^{+\infty}
\sum_{m=-\ell}^{+\ell} \psi_{\ell m} \left[2 \frac{\partial^2}{\partial \theta^2} +
\ell(\ell+1) \right]Y^{\ell m}, \label{hp_hc_a} \\
& & h_{\times}= \frac{1}{r} \sum_{\ell=2}^{+\infty}
\sum_{m=-\ell}^{+\ell} \psi_{\ell m}
\left[\frac{2}{\sin \theta}
\left(\frac{\partial^2}{\partial \theta \partial \varphi}-
\frac{\cos \theta}{\sin \theta}\frac{\partial}{\partial \varphi}
\right) \right]Y^{\ell m}. \nonumber \\
& & \label{hp_hc_b}
\end{eqnarray}
\end{subequations}
In these equations, we have introduced the gauge-invariant master functions $\psi_{\ell m} (t,r)$ of Cunningham, Price, and Moncrief \cite{Moncrief:1974am,Cunningham:1978zfa,Cunningham:1979px}. They can be written in the form \begin{equation}\label{TF_psi}
\psi_{\ell m} (t,r) = \frac{1}{\sqrt{2\pi}} \int_{-\infty}^{+\infty} d\omega \, \psi_{\omega \ell m} (r) e^{-i\omega t}
\end{equation}
where their Fourier components $\psi_{\omega \ell m} (r) $ satisfy the Zerilli-Moncrief equation (a radial Schr{\"o}dinger-like equation)
\begin{equation} \label{ZM EQ_Fourier}
\left[\frac{d^2}{d
r_\ast^2} + \omega^2 - V_\ell (r)  \right] \psi_{\omega \ell m} (r) = S_{\omega \ell m} (r).
\end{equation}
Here, $r_\ast$ denotes the tortoise coordinate which is defined in terms of the radial Schwarzschild coordinate $r$ by $dr/dr_\ast=f(r)$ and is given by $r_\ast(r)=r+2M \ln[r/(2M)-1]$ while $V_\ell (r)$ is the Zerilli-Moncrief potential given by
\begin{eqnarray} \label{pot Zerilli}
& & V_\ell(r)=f(r)\nonumber \\
& & \qquad \times \left[\frac{\Lambda^2(\Lambda+2) r^3+6\Lambda^2 Mr^2+36\Lambda M^2r+72M^3}{(\Lambda r+6M)^2r^3} \right]\nonumber \\
& &
\end{eqnarray}
with
\begin{equation} \label{Lambda_def}
\Lambda=(\ell -1) (\ell+2) = \ell (\ell+1)-2.
\end{equation}
The functions $S_{\omega \ell m} (r)$ appearing in the r.h.s.~of the Zerilli-Moncrief equation (\ref{ZM EQ_Fourier}) are source terms depending on the components, in the basis of tensor spherical harmonics, of the stress tensor inducing the perturbations of the Schwarzschild spacetime. Their general expression can be found in the review by Nagar and Rezzolla (cf.~Eq.~(4) of the erratum of Ref.~\cite{Nagar:2005ea}) and we have used it in connection with the geodesic equations (\ref{geodesic_equations}). We have obtained
\begin{equation} \label{SourceRad_gen}
S_{\omega \ell m} (r) =[Y^{\ell m}(\pi/2,0)]^\ast \widetilde{S}_{\omega \ell} (r) e^{+i \omega t_p(r) }
\end{equation}
where, for the particle starting at rest from infinity (i.e., for $\gamma=1$),
\begin{widetext}
\begin{subequations}  \label{SourceRad_omR}
\begin{eqnarray} \label{SourceRad_omR_a}
& & \widetilde{S}_{\omega \ell} (r) =\frac{8 \pi m_0}{\sqrt{2 \pi}(\Lambda+2)(\Lambda r + 6M)} f(r) \left[ -i \omega   \frac{r^2}{M}  +\sqrt{\frac{r}{2M}}\left(\frac{24M}{\Lambda r + 6M}-(\Lambda+1) \right) \right]
\end{eqnarray}
and
\begin{eqnarray}
\label{trajectory_Rad}
& & \frac{t_{p}(r)}{2M}=-\frac{2}{3} \left(\frac{r}{2M} \right)^{3/2} -2 \left(\frac{r}{2M} \right)^{1/2} + \ln \left(\frac{\sqrt{\frac{r}{2M}} +1}{\sqrt{\frac{r}{2M}} -1} \right) +  \frac{t_{0}}{2M},
\end{eqnarray}
\end{subequations}
and where, for a particle projected with a finite kinetic energy at infinity (i.e., for $\gamma > 1$),
\begin{subequations}\label{SourceRadRel_omR}
\begin{eqnarray} \label{SourceRadRel_omR_a}
& & \widetilde{S}_{\omega \ell} (r) =\frac{8 \pi m_0 }{\sqrt{2 \pi}(\Lambda+2)(\Lambda r + 6M)} f(r) \left[ -i \omega \frac{2\gamma  r^{2}}{(\gamma^{2}-1)r + 2M} \right.\nonumber \\
& &\left. -  \frac{\sqrt{r}\{\Lambda(\Lambda+2)(\gamma^{2}-1)r^{2}-2M\left[12(\gamma^{2}-1)^{2}-2(\Lambda-3)(\gamma^{2}-1)-\Lambda(\Lambda+1)\right] r- 12 M^{2}\left[5(\gamma^{2}-1)-(\Lambda-3)\right]\}}{(\Lambda r +6M)\left[(\gamma^{2}-1)r+2M\right]^{3/2}} \right] \nonumber \\
& &
\end{eqnarray}
and
\begin{eqnarray}
\label{trajectory_RadRel}
& & \frac{t_{p}(r)}{2M}=-\frac{\gamma}{(\gamma^{2}-1)^{3/2}}\sqrt{\left[(\gamma^{2}-1)\frac{r}{2M}\right]\left[(\gamma^{2}-1)\frac{r}{2M}+1\right]} \nonumber \\
& & - \frac{\gamma (2\gamma^{2}-3)}{(\gamma^{2}-1)^{3/2}} \ln\left[\sqrt{(\gamma^{2}-1)\frac{r}{2M}} + \sqrt{(\gamma^{2}-1)\frac{r}{2M}+1}\right] + \ln\left[\frac{\gamma \sqrt{\frac{r}{2M}}+ \sqrt{(\gamma^{2}-1)\frac{r}{2M}+1}}{ \gamma \sqrt{\frac{r}{2M}}- \sqrt{(\gamma^{2}-1)\frac{r}{2M}+1}}\right] +  \frac{t_{0}}{2M}.
\end{eqnarray}
\end{subequations}
In Eqs.~(\ref{trajectory_Rad}) and (\ref{trajectory_RadRel}), $t_0$ is an arbitrary integration constant.
\end{widetext}

\subsection{Zerilli-Moncrief equation and ${\cal S}$-matrix}
\label{SecIIb}

The Zerilli-Moncrief equation (\ref{ZM EQ_Fourier}) can be solved by using the machinery of Green's functions (see, e.g., Ref.~\cite{Breuer:1974uc} for its use in the context of BH physics). {\it Mutatis mutandis}, taking into account (\ref{SourceRad_gen}), the reasoning of Sec.~IIC of Ref.~\cite{Folacci:2018cic} permits us to obtain the asymptotic expression, for $r \to +\infty$, of the partial amplitudes $\psi_{\omega\ell m}(r)$. We have
\begin{subequations}\label{Partial_Response}
\begin{equation}
\label{Partial_Response_a}
\psi_{\omega\ell m}(r)= e^{+i \omega r_\ast(r) } \, \frac{K[\ell,\omega]}{2i\omega A^{(-)}_\ell (\omega)} \,
[Y^{\ell m}(\pi/2,0)]^\ast
\end{equation}
with
\begin{eqnarray}
\label{Partial_Response_b}
& & K[\ell,\omega] =  \int_{2M}^{+\infty} \frac{dr'}{f(r')} \,\phi_{\omega, \ell}^\mathrm {in}(r')
\, \widetilde{S}_{\omega \ell} (r') e^{i \omega t_p(r') }.
\end{eqnarray}
\end{subequations}
Here, we have introduced the solution $\phi_{\omega, \ell}^\mathrm {in} (r) $ of the homogeneous Zerilli-Moncrief equation
\begin{equation}
\label{H_ZM_equation}
\left[\frac{d^{2}}{dr_{\ast}^{2}}+\omega^{2}-V_{\ell}(r)\right] \phi_{\omega, \ell}^\mathrm {in}= 0
\end{equation}
which is defined by its behavior at the event horizon $r=2M$ (i.e., for $r_\ast \to -\infty$) and at spatial infinity $r \to +\infty$ (i.e., for $r_\ast \to +\infty$):
\begin{eqnarray}
\label{bc_in}
& & \phi_{\omega, \ell}^{\mathrm {in}}(r_{*}) \sim \left\{
\begin{aligned}
&\!\!\displaystyle{e^{-i\omega r_\ast}}  \,\, (r_\ast \to -\infty)\\
&\!\!\displaystyle{A^{(-)}_\ell (\omega) e^{-i\omega r_\ast} + A^{(+)}_\ell (\omega) e^{+i\omega r_\ast}} \,\, (r_\ast \to +\infty).
\end{aligned}
\right. \nonumber\\
&&
\end{eqnarray}
The coefficients $A^{(-)}_\ell (\omega)$ and  $A^{(+)}_\ell (\omega)$ appearing in Eqs.~(\ref{Partial_Response}) and (\ref{bc_in}), are complex amplitudes. By evaluating, first for $r_\ast \to - \infty$ and then for $r_\ast \to + \infty$, the Wronskian involving the function $\phi_{\omega\ell}^{\mathrm {in}}$ and its complex conjugate, we can show that they are linked by
\begin{equation}\label{Rel_conserv_Apm}
|A^{(-)}_\ell (\omega)|^2 - |A^{(+)}_\ell (\omega)|^2 = 1.
\end{equation}
Moreover, with the numerical calculation of the Weyl scalar $\Psi_4$ as well as the study of its properties in mind, it is important to note that
\begin{subequations}\label{Sym_om}
\begin{eqnarray}
& & \phi_{-\omega, \ell}^{\mathrm {in}} (r)=  \left[\phi_{\omega, \ell}^{\mathrm {in}}(r)  \right]^\ast, \label{Sym_om_a}\\
& & A^{(\pm )}_\ell (-\omega) =  [A^{( \pm)}_\ell (\omega) ]^\ast, \label{Sym_om_b}
\end{eqnarray}
and, as a consequence of the expressions (\ref{SourceRad_omR_a}) and (\ref{SourceRadRel_omR_a}) of the sources, that
\begin{equation}
K[\ell,-\omega] =  \left[ K[\ell,\omega] \right]^\ast \label{Sym_om_c}
\end{equation}
and
\begin{equation}
K[\ell,-\omega]/A^{(-)}_\ell (-\omega) =  \left[ K[\ell,\omega]/A^{(-)}_\ell (\omega) \right]^\ast. \label{Sym_om_d}
\end{equation}
\end{subequations}

It is worth pointing out that the expression (\ref{Partial_Response}) of the partial amplitudes $\psi_{\omega\ell m} (r)$ involves the ${\cal S}$-matrix defined by (see, e.g., Ref.~\cite{DeWitt:2003pm})
\begin{eqnarray}
\label{S_matrix_def}
& & {\cal S}_\ell (\omega) = \left(\,
\begin{aligned}
&\!\!\displaystyle{\qquad\quad 1/A^{(-)}_\ell (\omega)} &  \displaystyle{A^{(+)}_\ell (\omega)/A^{(-)}_\ell (\omega)} \\
&\!\!-\displaystyle{[A^{(+)}_\ell (\omega)]^\ast/A^{(-)}_\ell (\omega)} & \displaystyle{1/A^{(-)}_\ell (\omega)}\quad
\end{aligned}
\right). \nonumber\\
&&
\end{eqnarray}
We can note that, due to (\ref{Sym_om_b}), this matrix satisfies the symmetry property ${\cal S}_\ell (-\omega)=\left[ {\cal S}_\ell (\omega) \right]^\ast$ and that, due to (\ref{Rel_conserv_Apm}), it is in addition unitary, i.e., it satisfies $ {\cal S} {\cal S}^\dag= {\cal S}^\dag {\cal S}= \mathrm{1}$. Here, it is interesting to recall that, in Eq.~(\ref{S_matrix_def}), the term $1/A^{(-)}_\ell (\omega)$ and the term $A^{(+)}_\ell (\omega)/A^{(-)}_\ell (\omega)$ are, respectively, the transmission coefficient $T_\ell(\omega)$ and the reflection coefficient $R^\mathrm{in}_\ell(\omega)$ corresponding to the scattering problem defined by (\ref{bc_in}). As far as the coefficient $-[A^{(+)}_\ell (\omega)]^\ast/A^{(-)}_\ell (\omega)$ is concerned, it can be considered as the reflection coefficient $R^\mathrm{up}_\ell(\omega)$ involved in the scattering problem defining the modes $\phi^\mathrm{up}_{\omega, \ell} (r)$ \cite{DeWitt:2003pm}.

\subsection{Compact expression for the multipole expansion of the Weyl scalar $\Psi_4$}
\label{SecIIc}

We now substitute (\ref{TF_psi}) and (\ref{Partial_Response}) into (\ref{hp_hc_a}) and (\ref{hp_hc_b}). Furthermore, without loss of generality, we can assume that the gravitational wave is observed in a direction lying in the BH equatorial plane and making an angle $\varphi \in [0, \pi]$ with the trajectory of the particle (due to symmetry considerations, we can restrict our study to this interval). By then using the addition theorem for scalar spherical harmonics in the form
\begin{equation}\label{ThAd_HS}
\sum_{m=-\ell}^{+\ell}    Y^{\ell m}(\theta,\varphi) [Y^{\ell m}(\pi/2,0)]^\ast = \frac{2\ell +1}{4 \pi} P_\ell (\sin\theta \cos\varphi)
\end{equation}
where $P_\ell (x)$ denotes the Legendre polynomial of degree $\ell$ \cite{AS65}, we obtain, for $r\to +\infty$,
\begin{eqnarray}\label{hp_def}
& & r\, h_{+} (t,r,\theta=\pi/2,\varphi)= \frac{1}{\sqrt{2\pi}} \int_{-\infty}^{+\infty} d\omega \, e^{-i\omega[ t-r_\ast(r)]} \nonumber \\
& & \qquad\qquad \times \left[\sum_{\ell=2}^{+\infty} \frac{2\ell+1}{4\pi} \, \frac{K[\ell,\omega]}{2i\omega A^{(-)}_\ell (\omega)}  \, Z_\ell (\cos \varphi) \right]
\end{eqnarray}
and
\begin{equation}\label{hc_def}
r\, h_{\times} (t,r,\theta=\pi/2,\varphi)=0.
\end{equation}
In Eq.~(\ref{hp_def}), we have introduced the angular function
\begin{subequations}\label{AngFunction_Z}
\begin{eqnarray}\label{AngFunction_Za}
& & Z_\ell (\cos \varphi)= \left\{ \left[2 \frac{\partial^2}{\partial \theta^2} +
\ell(\ell+1) \right]P_\ell (\sin\theta \cos\varphi) \right\}_{\theta=\pi/2}
\end{eqnarray}
which, by using the properties of the Legendre polynomials \cite{AS65}, can be written in the form
\begin{eqnarray}\label{AngFunction_Zb}
& & Z_\ell (\cos \varphi)=  \frac{\ell +1}{\sin^2 \varphi} \left\{ 2\cos\varphi P_{\ell+1} (\cos\varphi)  \right. \nonumber \\
& & \qquad\qquad\quad \left. -\left[(\ell+2) \cos^2\varphi - \ell \right]P_\ell (\cos\varphi)  \right\}.
\end{eqnarray}
\end{subequations}

By inserting finally (\ref{hp_def}) and (\ref{hc_def}) into (\ref{Psi4_def}), we can write for $r \to + \infty$
\begin{eqnarray}\label{Psi4_ExpressionDef}
& & r\, \Psi_4 (t,r,\theta=\pi/2,\varphi)= \frac{1}{\sqrt{2\pi}} \int_{-\infty}^{+\infty} d\omega \, e^{-i\omega[ t-r_\ast(r)]} \nonumber \\
& & \qquad\qquad \times \left[ \sum_{\ell=2}^{+\infty} \frac{2\ell+1}{4\pi} \, \frac{i \omega K[\ell,\omega]}{4 A^{(-)}_\ell (\omega)}  \, Z_\ell (\cos \varphi) \right].
\end{eqnarray}
It should be noted that, due to the relation (\ref{Sym_om_d}), the term in squared brackets in the previous equation satisfies the Hermitian symmetry property and, as a consequence, the Weyl scalar $\Psi_4$ is a purely real quantity.

\subsection{Two alternative expressions for the multipole expansion of the Weyl scalar $\Psi_4$}
\label{SecIId}

It is very important to realize that (\ref{Psi4_ExpressionDef}) can also be written as
\begin{eqnarray}\label{Psi4_ExpressionDef_bis}
& & r\, \Psi_4 (t,r,\theta=\pi/2,\varphi)= \frac{1}{\sqrt{2\pi}} \int_{-\infty}^{+\infty} d\omega \, e^{-i\omega[ t-r_\ast(r)]} \nonumber \\
& & \qquad\qquad \times \left[ \sum_{\ell=0}^{+\infty} \frac{2\ell+1}{4\pi} \, \frac{i \omega K[\ell,\omega]}{4 A^{(-)}_\ell (\omega)}  \, Z_\ell (\cos \varphi) \right].
\end{eqnarray}
Indeed, it is possible to start at $\ell=0$ the discrete sum over $\ell$ taking into account the results
\begin{equation}\label{prop_Zell}
Z_0 (\cos \varphi)=0 \quad \mathrm{and} \quad Z_1 (\cos \varphi)=0
\end{equation}
which are easily obtained from the definition (\ref{AngFunction_Z}) by noting that $P_0(x)=1$ and $P_1(x)=x$. Of course, in general, it is more natural to work with the multipole expansion (\ref{Psi4_ExpressionDef}) of the Weyl scalar $\Psi_4$ but, in Sec.~\ref{SecIIIc}, we will take (\ref{Psi4_ExpressionDef_bis}) as a departure point because it will permit us to use the Poisson summation formula in its standard form.

It is moreover interesting to note that (\ref{Psi4_ExpressionDef_bis}) can be rewritten in the form
\begin{eqnarray}\label{Psi4_ExpressionDef_ter}
& & r\, \Psi_4 (t,r,\theta=\pi/2,\varphi)= \frac{1}{\sqrt{2\pi}} \int_{-\infty}^{+\infty} d\omega \, e^{-i\omega[ t-r_\ast(r)]} \nonumber \\
& & \qquad \times \left[ \sum_{\ell=0}^{+\infty} (-1)^\ell \frac{2\ell+1}{4\pi} \, \frac{i \omega K[\ell,\omega]}{4 A^{(-)}_\ell (\omega)}  \, Z_\ell (-\cos \varphi) \right].
\end{eqnarray}
Indeed, we can recover (\ref{Psi4_ExpressionDef_bis}) from (\ref{Psi4_ExpressionDef_ter}) by using
\begin{equation}\label{prop_Zell_lml}
Z_\ell (- \cos \varphi) = (-1)^\ell Z_\ell (\cos \varphi)
\end{equation}
which is a direct consequence of the definition (\ref{AngFunction_Z}), of the properties of the Legendre polynomials and, in particular, of the relation \cite{AS65}
\begin{equation}\label{prop_Pell}
P_\ell (- \cos \varphi) = (-1)^\ell P_\ell (\cos \varphi).
\end{equation}
In Sec.~\ref{SecIIId}, we will take (\ref{Psi4_ExpressionDef_ter}) as a departure point because it will permit us to use the Sommerfeld-Watson transform in its standard form.

\subsection{Quasinormal ringdown associated with the Weyl scalar $\Psi_4$}
\label{SecIIe}

We can extract from (\ref{Psi4_ExpressionDef}) the quasinormal ringdown $\Psi^\text{\tiny{QNM}}_4$ generated by the massive particle falling radially from infinity into a Schwarzschild BH. This is achieved by deforming the contour of integration over $\omega$ in Eq.~(\ref{Psi4_ExpressionDef}) (see, e.g., Ref.~\cite{Leaver:1986gd}). This deformation permits us to capture the zeros of the coefficients $A^{(-)}_\ell (\omega)$ lying in the lower part of the complex $\omega$ plane, i.e., the solutions of the equation
\begin{equation}\label{freqQN_def_Am}
A^{(-)}_{\ell} (\omega_{\ell n})=0
\end{equation}
where $\omega_{\ell n}$ denotes the complex frequencies of the $(\ell,n)$ QNMs. We recall that the quasinormal-frequency spectrum is symmetric with respect to the imaginary axis, i.e., if $\omega_{\ell n}$ is a quasinormal frequency lying in the fourth quadrant, $-\omega_{\ell n}^{*}$ is the symmetric quasinormal frequency lying in the third one. We also recall that, for a given $\ell$, $n=1$ corresponds to the fundamental QNM (i.e., the least damped one) and $n=2, 3, \dots$ correspond to the overtones. By using Cauchy's theorem and introducing the residues of the $\cal{S}$-matrix ${\cal S}_\ell (\omega)$ [or, more precisely, of the function $1/A^{(-)}_\ell (\omega)$] at $\omega=\omega_{\ell n}$ and $\omega=-\omega_{\ell n}^{*}$, we then easily obtain
 \begin{eqnarray}
\label{response_QNM}
& & r\, \Psi^\text{\tiny{QNM}}_4 (t,r,\theta=\pi/2,\varphi)=  \sqrt{2 \pi} \, \operatorname{Re}   \left[ \, \sum^{+\infty}_{\ell =2} \sum^{+\infty}_{n =1} \phantom{\frac{K[\ell,\omega_{\ell n}]}{A_{\ell}^{(+)}(\omega_{\ell n})}}  \right. \nonumber \\
& & \,\, \left. \frac{2\ell +1}{4\pi} {\cal{B}}_{\ell n} \frac{(\omega_{\ell n})^2 K[\ell,\omega_{\ell n}]}{A_{\ell}^{(+)}(\omega_{\ell n})}\, e^{-i \omega_{\ell n}[t-r_\ast(r)]}  \, Z_\ell (\cos \varphi)\right]. \nonumber\\
& &
\end{eqnarray}
In the previous expression,
\begin{equation}
\label{excitation_factor_QNM}
{\cal{B}}_{\ell n}=\left[\frac{1}{2 \omega}\,\,\frac{A_{\ell}^{(+)}(\omega)}{\frac{d}{d \omega}A_{\ell}^{(-)}(\omega)}\right]_{\omega=\omega_{\ell n}}
\end{equation}
denotes the excitation factor associated with the $(\ell,n)$ QNM of complex frequency $\omega_{\ell n}$. Its expression involves the residue of the function $1/A^{(-)}_\ell (\omega)$ at $\omega=\omega_{\ell n}$. It should be noted that, in order to obtain (\ref{response_QNM}), we have gathered the contributions of the quasinormal frequencies $\omega_{\ell n}$ and $-\omega_{\ell n}^\ast$ taking into account the relations (\ref{Sym_om_b}) and (\ref{Sym_om_c}) which remain valid in the complex $\omega$ plane. As a consequence, the quasinormal ringdown waveform $\Psi^\text{\tiny{QNM}}_4$ appears clearly as a purely real quantity.

It is important to recall that the ringdown waveform $\Psi^\text{\tiny{QNM}}_4$ does not provide physically relevant results at early times due to the exponentially divergent behavior of each of its components as $t$ decreases. It is therefore necessary to determine, from physical considerations (see below), a starting time $t_\mathrm{start}$ for the BH ringdown.

\section{The Weyl scalar $\Psi_4$, its CAM representations and its Regge pole approximations}
\label{SecIII}

In this section, we shall provide two CAM representations of the Weyl scalar $\Psi_4$. There are exact representations that are obtained by replacing the discrete sum over integer values of the angular momentum $\ell$ by a sum over Regge poles plus background integrals along the positive real axis and the imaginary axis of the CAM plane. We shall also discuss the Regge pole part of these representations as approximations of the Weyl scalar $\Psi_4$.

\subsection{Some preliminary remarks concerning analytic extensions in the CAM plane}
\label{SecIIIa}

A CAM representation of the multipolar waveform $\Psi_4$ given by Eq.~(\ref{Psi4_ExpressionDef}) can be obtained by first applying the Poisson summation formula \cite{MorseFeshbach1953} to (\ref{Psi4_ExpressionDef_bis}) or, equivalently, the Sommerfeld-Watson transform \cite{Watson18,Sommerfeld49,Newton:1982qc} to (\ref{Psi4_ExpressionDef_ter}), and then by using Cauchy's theorem. As we shall see below, this requires to replace, into the term
\begin{equation} \label{from_ell_to_lambda_Poisson}
\sum_{\ell=0}^{+\infty} \frac{2\ell+1}{4\pi} \, \frac{i \omega K[\ell,\omega]}{4 A^{(-)}_\ell (\omega)}  \, Z_\ell (\cos \varphi)
\end{equation}
of Eq.~(\ref{Psi4_ExpressionDef_bis}), or into the term
\begin{equation} \label{from_ell_to_lambda_SWT}
\sum_{\ell=0}^{+\infty} (-1)^\ell \frac{2\ell+1}{4\pi} \, \frac{i \omega K[\ell,\omega]}{4 A^{(-)}_\ell (\omega)}  \, Z_\ell (-\cos \varphi)
\end{equation}
of Eq.~(\ref{Psi4_ExpressionDef_ter}), the angular momentum $\ell \in \mathbb{N}$ by the angular momentum $\lambda = \ell +1/2 \in \mathbb{C}$ and to work into the CAM plane. As a consequence, we need to have at our disposal the functions $Z_{\lambda -1/2} (\cos \varphi)$, $Z_{\lambda -1/2} (-\cos \varphi)$ and $K[\lambda-1/2,\omega]/A^{(-)}_{\lambda-1/2} (\omega)$ which are ``the'' analytic extensions of $Z_{\ell} (\cos \varphi)$, $Z_{\ell} (-\cos \varphi)$ and $K[\ell,\omega]/A^{(-)}_{\ell} (\omega)$ in the complex $\lambda$ plane. In fact, the uniqueness of these analytic extensions is a difficult mathematical problem that goes beyond the scope of our study. In general (i.e., in the context of the resolution of a Schr{\"o}dinger-like equation with an arbitrary potential and of the CAM analysis of the associated resonant scattering), as noted by Newton in Ch.~13 of Ref.~\cite{Newton:1982qc}, there are infinitely many ways of constructing an analytic function taking prescribed values for integers and the ``right'' one is justified by the results.

It should be noted, however, that, as far as the determination of the analytic extension of $Z_{\ell} (\cos \varphi)$ and $Z_{\ell} (-\cos \varphi)$ is concerned, the problem can be easily solved. Indeed, these two angular functions can be expressed in terms of Legendre polynomials [see Eq.~(\ref{AngFunction_Z})] of which the analytic extension has been widely discussed in the literature concerning the CAM approach of resonant scattering. We recall that the analytic extension of $P_\ell (z)$ usually considered is the hypergeometric function \cite{AS65}
\begin{equation}\label{Def_ext_LegendreP}
P_{\lambda -1/2} (z) = F(1/2-\lambda,1/2+\lambda;1;(1-z)/2]
\end{equation}
and it is worth noting that it satisfies
\begin{equation}\label{prop_ext_LegendreP}
P_{-\lambda -1/2} (z) = P_{\lambda -1/2} (z).
\end{equation}
From (\ref{AngFunction_Z}), we then can write
\begin{subequations}\label{AngFunction_Z_ext}
\begin{eqnarray}\label{AngFunction_Za_ext}
& & Z_{\lambda -1/2} (\pm \cos \varphi)= \left\{ \left[2 \frac{\partial^2}{\partial \theta^2} +
(\lambda^2 -1/4) \right] \right. \nonumber \\
& & \qquad\qquad \left.  \phantom{ \left[2 \frac{\partial^2}{\partial \theta^2} \right]} P_{\lambda -1/2} (\pm \sin\theta \cos\varphi) \right\}_{\theta=\pi/2}
\end{eqnarray}
from which we obtain
\begin{eqnarray}\label{AngFunction_Zb_ext}
& & Z_{\lambda -1/2} (\pm \cos \varphi)=  \frac{\lambda +1/2}{\sin^2 \varphi} \left\{ \pm 2\cos\varphi P_{\lambda +1/2} (\pm \cos\varphi)  \right. \nonumber \\
& & \quad \left. -\left[(\lambda +3/2) \cos^2\varphi - (\lambda -1/2) \right]P_{\lambda -1/2} (\pm \cos\varphi)  \right\}. \nonumber \\
& &
\end{eqnarray}
\end{subequations}
We can immediately check that, due to (\ref{Def_ext_LegendreP}) and (\ref{prop_ext_LegendreP}), we have
\begin{equation}\label{prop_ext_Z_a}
Z_{-\lambda -1/2} (\pm \cos \varphi) = Z_{\lambda -1/2} (\pm \cos \varphi)
\end{equation}
and
\begin{equation}\label{prop_ext_Z_b}
Z_{\lambda -1/2} (\pm \cos \varphi) = [Z_{\lambda^\ast -1/2} (\pm \cos \varphi)]^\ast.
\end{equation}
Here, it is very important to note that, while the angular functions $Z_\ell (\pm \cos \phi)$ are well defined for $\varphi \in [0,\pi]$, this is not the case for their analytic extensions $Z_{\lambda -1/2} (\pm \cos \phi)$. Indeed, due to the pathologic behavior of $P_{\lambda -1/2} (z)$ at $z=-1$ [see, Eq.~(\ref{Def_ext_LegendreP})], $Z_{\lambda -1/2} (\cos \phi)$ diverges in the limit $\varphi \to \pi$ and $Z_{\lambda -1/2} (-\cos \phi)$ diverges in the limit $\varphi \to 0$. We shall return to these results later due to the problems they generate on the Regge pole approximation of $\Psi_4$.

An analytic extension $K[\lambda-1/2,\omega]/A^{(-)}_{\lambda-1/2} (\omega)$ of $K[\ell,\omega]/A^{(-)}_{\ell} (\omega)$ must satisfy a generalization of the relation (\ref{Sym_om_d}). We first note that the function $\phi_{\omega, \lambda-1/2}^\mathrm {in} (r) $ and the coefficients $A^{(\pm)}_{\lambda-1/2} (\omega)$ that are defined by the problem (\ref{H_ZM_equation})-(\ref{bc_in}) where now $\ell \in \mathbb{N}$ is replaced by $\lambda-1/2 \in \mathbb{C}$, satisfy
\begin{subequations}\label{Sym_om_CAM}
\begin{eqnarray}
& & \phi_{-\omega, \lambda -1/2}^{\mathrm {in}} (r)=  [\phi_{\omega, \lambda^\ast -1/2}^{\mathrm {in}}(r)]^\ast, \label{Sym_om_CAM_a}\\
& & A^{(\pm )}_{\lambda -1/2} (-\omega) =  [A^{( \pm)}_{\lambda^\ast -1/2} (\omega) ]^\ast. \label{Sym_om_CAM_b}
\end{eqnarray}
We also note that, as a consequence of the expressions (\ref{SourceRad_omR_a}) and (\ref{SourceRadRel_omR_a}) of the sources, we have
\begin{equation}
K[\lambda -1/2,-\omega] =  \left[ K[\lambda^\ast -1/2,\omega] \right]^\ast \label{Sym_om_CAM_c}
\end{equation}
and therefore
\begin{eqnarray}\label{Sym_om_CAM_d}
& & K[\lambda -1/2,-\omega]/A^{(-)}_{\lambda -1/2} (-\omega) =  \nonumber \\
& & \qquad\qquad\qquad \left[ K[\lambda^\ast -1/2,\omega]/A^{(-)}_{\lambda^\ast -1/2} (\omega) \right]^\ast.
\end{eqnarray}
\end{subequations}

\begin{figure*}
\centering
 \includegraphics[scale=0.55]{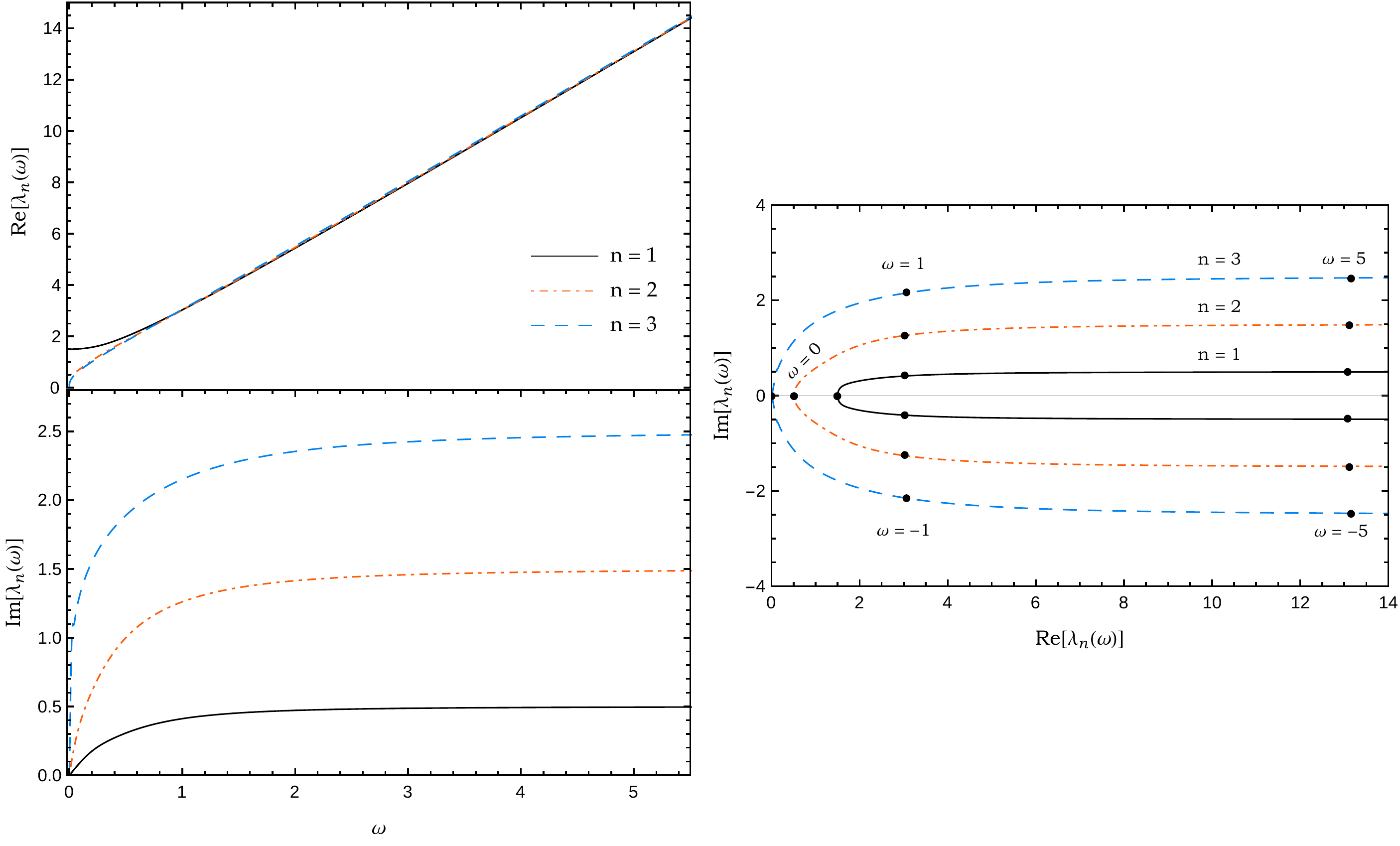}
\caption{\label{TR_PR_S_2} Regge trajectories of the first three Regge poles corresponding to the even-parity perturbations of the Schwarzschild BH ($2M=1$). The relation (\ref{PR_chgt_om}) permits us to describe the Regge trajectories for $\omega<0$ by noting that $\operatorname{Re} [\lambda_n(\omega)]$ and $\operatorname{Im} [\lambda_n(\omega)]$ are, respectively, even and odd functions of $\omega$. We observe, in particular, the migration of the Regge poles in the CAM plane.}
\end{figure*}

\begin{figure*}
\centering
 \includegraphics[scale=0.55]{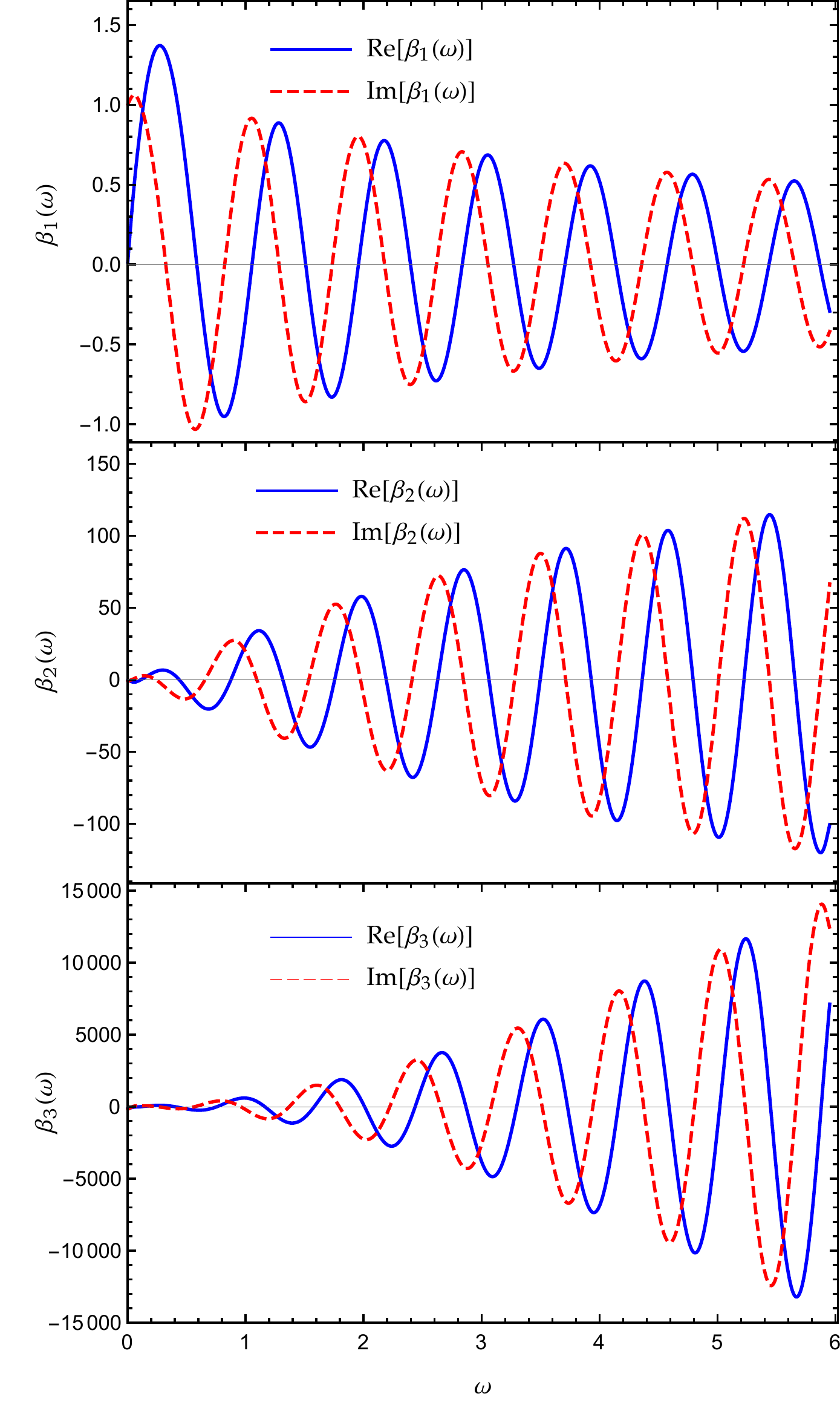}
\caption{\label{RM_excitation_factors_even} Regge trajectories of the Regge-mode excitation factors ($2M=1$). We consider the Regge modes corresponding to the first three Regge poles of which the behavior has been displayed in Fig.~\ref{TR_PR_S_2} ($2M=1$). The relation (\ref{EF_PR_chgt_om}) permits us to describe the Regge trajectories for $\omega<0$ by noting that $\operatorname{Re} [\beta_n(\omega)]$ and $\operatorname{Im} [\beta_n(\omega)]$ are, respectively, odd and even functions of $\omega$.}
\end{figure*}

\subsection{Regge poles, Regge modes and associated excitation factors}
\label{SecIIIb}

In the next two subsections, contour deformations in the CAM plane will permit us to collect, by using Cauchy's residue theorem, the contributions from the Regge poles of the ${\cal S}$-matrix or, more precisely, from the poles, in the complex $\lambda$ plane and for $\omega \in \mathbb{R}$, of the matrix ${\cal S}_{\lambda-1/2}(\omega)$. It should be noted that these poles can be defined as the zeros $\lambda_n(\omega)$ with $n=1, 2, 3, \dots  $ and $\omega \in \mathbb{R}$ of the coefficient $A^{(-)}_{\lambda-1/2} (\omega)$ [see Eq.~(\ref{S_matrix_def})]. They therefore satisfy
\begin{equation}\label{PR_def_Am}
A^{(-)}_{\lambda_n(\omega)-1/2} (\omega)=0.
\end{equation}

The Regge poles corresponding to the odd-parity perturbations of the Schwarzschild BH have been studied in Refs.~\cite{Decanini:2002ha,Decanini:2009mu,Dolan:2009nk}. But, here, we have to consider the Regge poles corresponding to the even-parity perturbations. While the odd-parity sector is governed by the Regge-Wheeler equation, the even-parity one is governed by the Zerilli-Moncrief equation and, as a consequence, the ${\cal S}$-matrix differs according to the parity sector. However, it is important to recall that the solutions of the homogeneous Zerilli-Moncrief and Regge-Wheeler equations are related by the Chandrasekhar-Detweiler transformation \cite{Chandrasekhar:1975zza,Chandrasekhar:1985kt}. As a consequence, by using the fact that the coefficients $A^{(-)}_\ell (\omega)$ and therefore $A^{(-)}_{\lambda-1/2} (\omega)$ are identical in the two parity sectors, it is obvious that the Regge pole spectrum does not depend on the parity sector and the results already obtained for the Regge poles corresponding to the odd-parity perturbations of the Schwarzschild BH can be used here without any change. In particular, it should be recalled that, for $\omega >0$, the Regge poles lie in the first and third quadrants of the CAM plane, symmetrically distributed with respect to the origin $O$ of this plane. In this article, due to the use of Fourier transforms, we must be able to locate the Regge poles even for $\omega < 0$. In fact, from the symmetry relation (\ref{Sym_om_CAM_b}), we have
\begin{equation}\label{PR_chgt_om}
\lambda_n(-\omega)=[\lambda_n(\omega)]^\ast
\end{equation}
and we can see immediately that, for $\omega < 0$, the Regge poles lie in the second and fourth quadrants of the CAM plane, symmetrically distributed with respect to the origin $O$ of this plane. Moreover, if we consider the Regge trajectories $\lambda_n(\omega)$ with $\omega \in ]-\infty,+\infty[$, we can observe the migration of the Regge poles. More precisely, as $\omega $ decreases, the Regge poles lying in the first (third) quadrant of the CAM plane migrate in the fourth (second) one.

It should be noted that the solutions of the problem (\ref{H_ZM_equation})-(\ref{bc_in}) with $\ell$ replaced by $\lambda_n(\omega)-1/2 $ are modes that are purely outgoing at infinity and purely ingoing at the horizon. They are the ``Regge modes'' of the Schwarzschild BH \cite{Decanini:2009mu}. Because of the analogy with the QNMs, it is natural to define excitation factors for these modes. In fact, they will appear in the CAM representation of the Weyl scalar $\Psi_4$. By analogy with the excitation factor associated with the $(\ell,n)$ QNM of complex frequency $\omega_{\ell n}$ [see Eq.~(\ref{excitation_factor_QNM})], we define the excitation factor of the Regge mode associated with the Regge pole $\lambda_n(\omega)$ by
\begin{equation}\label{excitation_factor_RP}
\beta_n(\omega)=\left[\frac{1}{2 \omega}\,\,\frac{A_{\lambda -1/2}^{(+)}(\omega)}{\frac{d}{d \lambda}A_{\lambda -1/2}^{(-)}(\omega)}\right]_{\lambda=\lambda_n(\omega)}.
\end{equation}
Its expression involves the residue of the matrix ${\cal S}_{\lambda-1/2}(\omega)$ [or, more precisely, of the function $1/A^{(-)}_{\lambda-1/2} (\omega)$] at $\lambda=\lambda_n(\omega)$. It should be noted that, due to (\ref{Sym_om_CAM_b}), we have
\begin{equation}\label{EF_PR_chgt_om}
\beta_n(-\omega)=-[\beta_n(\omega)]^\ast.
\end{equation}
It is worth pointing out that, unlike the Regge poles, the excitation factors (\ref{excitation_factor_RP}) depend on the parity sector because the coefficients $A^{(+)}_\ell (\omega)$ and therefore $A^{(+)}_{\lambda-1/2} (\omega)$ are parity dependent \cite{Chandrasekhar:1975zza,Chandrasekhar:1985kt}.

We have displayed the Regge trajectories of the first three Regge poles as well as the Regge trajectories of the corresponding excitation factors in Figs.~\ref{TR_PR_S_2} and \ref{RM_excitation_factors_even}. These numerical results have been obtained by using, {\it mutatis mutandis}, the methods that have permitted us to obtain, in Refs.~\cite{Folacci:2018vtf} and \cite{Folacci:2018cic}, for the electromagnetic field and for the gravitational waves, the complex quasinormal frequencies of the QNMs and the associated excitation factors (see, e.g., Sec.~IVA of Ref.~\cite{Folacci:2018cic}). The numerical calculations have been performed using {\it Mathematica} \cite{Mathematica}.

It is important to recall that, in Refs.~\cite{Decanini:2002ha,Decanini:2009mu}, we have established a connection between the Regge modes and the (weakly damped) QNMs of the Schwarzschild BH. It will play a central role in the interpretation of our results in Sec.~\ref{SecIV} and in the Conclusion, and we recall that, for a given $n$, the Regge trajectory $\lambda_n(\omega)$ with $\omega \in \mathbb{R}$ encodes information on the complex quasinormal frequencies $\omega_{\ell n}$ with $\ell=2, 3, \dots$ In fact, the index $n=1, 2, 3, \dots  $ permits us not only to distinguish between the different Regge poles but is also associated with the family of quasinormal frequencies generated by the Regge modes.

\subsection{CAM representation and Regge pole approximation of the Weyl scalar $\Psi_4$ based on the Poisson summation formula}
\label{SecIIIc}

By means of the usual ``half-range'' Poisson summation formula \cite{MorseFeshbach1953}
\begin{equation}\label{HR_Poisson}
\sum_{\ell=0}^{+\infty} F(\ell+1/2)=\sum_{p=-\infty}^{+\infty}
(-1)^p \int_0^{+\infty} d\lambda \, F(\lambda)e^{i 2 \pi p \lambda}
\end{equation}
applied, in Eq.~(\ref{Psi4_ExpressionDef_bis}), to the discrete sum over the ordinary angular momentum $\ell$, we obtain
\begin{widetext}
\begin{eqnarray}\label{PSF_Psi4_ExpressionDef}
& & r\, \Psi_4 (t,r,\theta=\pi/2,\varphi)= \frac{1}{\sqrt{2\pi}} \int_{-\infty}^{+\infty} d\omega \, e^{-i\omega[ t-r_\ast(r)]} \left[ \int_{0}^{+\infty} d\lambda \, \frac{\lambda}{2\pi}  \frac{i \omega K[\lambda-1/2,\omega]}{4 A^{(-)}_{\lambda-1/2} (\omega)}  \, Z_{\lambda -1/2} (\cos \varphi) \right. \nonumber \\
& & \qquad \qquad\qquad \left. + \sum_{p=1}^{+\infty} \int_{0}^{+\infty} d\lambda \, \frac{\lambda}{2\pi}  \frac{i \omega K[\lambda-1/2,\omega])}{4 A^{(-)}_{\lambda-1/2} (\omega)}  \, Z_{\lambda -1/2} (\cos \varphi) e^{i 2 \pi p(\lambda-1/2)} \right.  \nonumber \\
& & \qquad \qquad\qquad \left.  + \sum_{p=1}^{+\infty} \int_{0}^{+\infty} d\lambda \, \frac{\lambda}{2\pi}  \frac{i \omega K[\lambda-1/2,\omega]}{4 A^{(-)}_{\lambda-1/2} (\omega)}  \, Z_{\lambda -1/2} (\cos \varphi) e^{-i 2 \pi p(\lambda-1/2)}     \right].
\end{eqnarray}
\end{widetext}
The integrals in the second term of the r.h.s.~of (\ref{PSF_Psi4_ExpressionDef}) can
be evaluated by using Cauchy's residue theorem. This is achieved by closing the path
along the positive real axis with a quarter circle at infinity in
the first quadrant of the CAM plane and a path
along the positive imaginary axis going from $+i\infty$ to $0$. The
integrals in the third term of the r.h.s.~of (\ref{PSF_Psi4_ExpressionDef}) can be
evaluated similarly but now by closing the path along the positive
real axis in the fourth quadrant of the CAM
plane. Cauchy's residue theorem permits us to collect the contributions from the Regge poles $\lambda_n(\omega)$ with $n=1, 2, 3, \dots  $, i.e., the zeros for $\omega \in \mathbb{R}$ of the coefficient $A^{(-)}_{\lambda-1/2} (\omega)$. Here, it should be recalled that these poles of the ${\cal S}$-matrix lie in the first quadrant of the CAM plane for $\omega >0$ and that they migrate in the fourth one for $\omega <0$ (see also Sec.~\ref{SecIIIb}). By assuming that the contributions coming from the two quarter circles at infinity vanish [a drastic assumption that implies strong constraints on the choice of the analytic extension of $K[\ell,\omega]/A^{(-)}_{\ell} (\omega)$ in the complex $\lambda$ plane], we then obtain
\begin{widetext}
\begin{eqnarray}\label{PSF_Psi4_ExpressionDef_v2}
& & r\, \Psi_4 (t,r,\theta=\pi/2,\varphi)= \frac{1}{\sqrt{2\pi}} \int_{-\infty}^{+\infty} d\omega \, e^{-i\omega[ t-r_\ast(r)]} \left[ \int_{0}^{\infty} d\lambda \, \frac{\lambda}{2\pi}  \frac{i \omega K[\lambda-1/2,\omega]}{4 A^{(-)}_{\lambda-1/2} (\omega)}  \, Z_{\lambda -1/2} (\cos \varphi) \right. \nonumber \\
& & \qquad \qquad\qquad \left. + \sum_{p=1}^{+\infty} \int_{0}^{+i\infty} d\lambda \, \frac{\lambda}{2\pi}  \frac{i \omega K[\lambda-1/2,\omega]}{4 A^{(-)}_{\lambda-1/2} (\omega)}  \, Z_{\lambda -1/2} (\cos \varphi) e^{i 2 \pi p(\lambda-1/2)} \right.  \nonumber \\
& & \qquad \qquad\qquad \left.  + \sum_{p=1}^{+\infty} \int_{0}^{-i\infty} d\lambda \, \frac{\lambda}{2\pi}  \frac{i \omega K[\lambda-1/2,\omega]}{4 A^{(-)}_{\lambda-1/2} (\omega)}  \, Z_{\lambda -1/2} (\cos \varphi) e^{-i 2 \pi p(\lambda-1/2)}  \right. \nonumber \\
& & \qquad \qquad\qquad \left. - {\cal H}(\omega)\sum_{p=1}^{+\infty} \sum_{n=1}^{+\infty}  \lambda_n(\omega) \beta_n(\omega) \, \frac{\omega^2 K[\lambda_n(\omega)-1/2,\omega]}{ 2 A^{(+)}_{\lambda_n(\omega)-1/2} (\omega)} \,  Z_{\lambda_n(\omega) -1/2} (\cos \varphi) e^{i 2 \pi p(\lambda_n(\omega)-1/2)} \right. \nonumber \\
& & \qquad \qquad\qquad \left. + {\cal H}(-\omega)\sum_{p=1}^{+\infty} \sum_{n=1}^{+\infty}  \lambda_n(\omega) \beta_n(\omega) \, \frac{\omega^2 K[\lambda_n(\omega)-1/2,\omega]}{ 2 A^{(+)}_{\lambda_n(\omega)-1/2} (\omega)} \,  Z_{\lambda_n(\omega) -1/2} (\cos \varphi) e^{-i 2 \pi p(\lambda_n(\omega)-1/2)}  \right]
\end{eqnarray}
\end{widetext}
where ${\cal H}$ denotes the Heaviside step function. In Eq.~(\ref{PSF_Psi4_ExpressionDef_v2}), we have introduced the excitation factor (\ref{excitation_factor_RP}) of the Regge mode associated with the Regge pole $\lambda_n(\omega)$. We can now simplify (\ref{PSF_Psi4_ExpressionDef_v2}) by
using the relations
\begin{subequations}\label{devS sin}
\begin{eqnarray}
& &\sum_{p=1}^{+\infty} e^{i 2\pi p(z -1/2)}=- \frac{e^{i
\pi z}}{2 \cos (\pi z) } \quad \mathrm{valid \,\, if} \,\,
\operatorname{Im} \, z > 0, \label{devS sin_a} \nonumber \\
& &   \\
& &\sum_{p=1}^{+\infty} e^{-i 2 \pi p(z -1/2)}=
-\frac{e^{-i
\pi z}}{2 \cos (\pi z)} \quad \mathrm{valid \,\, if} \,\,
\operatorname{Im} \ z < 0. \label{devS sin_b} \nonumber \\
& &
\end{eqnarray}
\end{subequations}
We then obtain
\begin{widetext}
\begin{equation}\label{CAM_Psi4_ExpressionDef_Ptot}
\Psi_4 (t,r,\theta=\pi/2,\varphi)= \Psi^{\text{\tiny{B}} \, \textit{\tiny{(P)}}}_4 (t,r,\theta=\pi/2,\varphi) + \Psi^{\text{\tiny{RP}} \, \textit{\tiny{(P)}}}_4 (t,r,\theta=\pi/2,\varphi)
\end{equation}
where
\begin{subequations}\label{CAM_Psi4_ExpressionDef_P}
\begin{eqnarray}\label{CAM_Psi4_ExpressionDef_P_Background}
& & r\, \Psi^{\text{\tiny{B}} \, \textit{\tiny{(P)}}}_4 (t,r,\theta=\pi/2,\varphi)= \frac{1}{\sqrt{2\pi}} \int_{-\infty}^{+\infty} d\omega \, e^{-i\omega[ t-r_\ast(r)]} \left[ \int_{0}^{\infty} d\lambda \, \frac{\lambda}{2\pi}  \frac{i \omega K[\lambda-1/2,\omega]}{4 A^{(-)}_{\lambda-1/2} (\omega)}  \, Z_{\lambda -1/2} (\cos \varphi) \right. \nonumber \\
& & \qquad \qquad\qquad \left. -\frac{1}{4\pi} \int_{0}^{+i\infty} d\lambda \, \frac{\lambda e^{i\pi \lambda}}{\cos (\pi \lambda)}  \frac{i \omega K[\lambda-1/2,\omega]}{4 A^{(-)}_{\lambda-1/2} (\omega)}  \, Z_{\lambda -1/2} (\cos \varphi)   \right.  \nonumber \\
& & \qquad \qquad\qquad \left.  -\frac{1}{4\pi} \int_{0}^{-i\infty} d\lambda \, \frac{\lambda e^{-i \pi \lambda}}{\cos (\pi \lambda)}  \frac{i \omega K[\lambda-1/2,\omega]}{4 A^{(-)}_{\lambda-1/2} (\omega)}  \, Z_{\lambda -1/2} (\cos \varphi)   \right]
\end{eqnarray}
is a background integral contribution and where
\begin{eqnarray}\label{CAM_Psi4_ExpressionDef_P_RP}
& & r\, \Psi^{\text{\tiny{RP}} \, \textit{\tiny{(P)}}}_4 (t,r,\theta=\pi/2,\varphi)=\frac{1}{\sqrt{2\pi}} \int_{-\infty}^{+\infty} d\omega \, e^{-i\omega[ t-r_\ast(r)]} \nonumber \\
& & \qquad \left[  {\cal H}(\omega)  \sum_{n=1}^{+\infty}  \frac{\lambda_n(\omega) \beta_n(\omega) e^{i\pi \lambda_n(\omega)}}{\cos [\pi \lambda_n(\omega)]} \, \frac{\omega^2 K[\lambda_n(\omega)-1/2,\omega]}{ 4 A^{(+)}_{\lambda_n(\omega)-1/2} (\omega)} \,  Z_{\lambda_n(\omega) -1/2} (\cos \varphi)   \right. \nonumber \\
& & \qquad \qquad\qquad \left. - {\cal H}(-\omega) \sum_{n=1}^{+\infty}  \frac{\lambda_n(\omega) \beta_n(\omega) e^{-i\pi \lambda_n(\omega)}}{\cos [\pi \lambda_n(\omega)]} \, \, \frac{\omega^2 K[\lambda_n(\omega)-1/2,\omega]}{ 4 A^{(+)}_{\lambda_n(\omega)-1/2} (\omega)} \,  Z_{\lambda_n(\omega) -1/2} (\cos \varphi)    \right]
\end{eqnarray}
\end{subequations}
\end{widetext}
is the Fourier transform of a sum over Regge poles. We can again check that $\Psi_4$ is a real-valued  function by now considering this new expression. Indeed, due to the relations (\ref{prop_ext_Z_b}) and (\ref{Sym_om_CAM_d}), the first term as well as the sum of the second and third terms into the squared bracket in the r.h.s.~of (\ref{CAM_Psi4_ExpressionDef_P_Background}) satisfy the Hermitian symmetry property. Such a property is also satisfied by the sum of the two terms into the squared bracket in the r.h.s.~of (\ref{CAM_Psi4_ExpressionDef_P_RP}) as a consequence of the relations (\ref{prop_ext_Z_b}), (\ref{Sym_om_CAM_b}), (\ref{Sym_om_CAM_c}), (\ref{PR_chgt_om}) and (\ref{EF_PR_chgt_om}).

It is important to note that (\ref{CAM_Psi4_ExpressionDef_Ptot}) provides an exact expression for the Weyl scalar $\Psi_4$, equivalent to the initial expression (\ref{Psi4_ExpressionDef}). From this CAM representation of $\Psi_4$, we can extract the contribution denoted by $\Psi^{\text{\tiny{RP}} \, \textit{\tiny{(P)}}}_4$ and given by (\ref{CAM_Psi4_ExpressionDef_P_RP}) which, as a sum over Regge poles, is only an approximation of $\Psi_4$. In Sec.~\ref{SecIV}, we shall compare it with the exact expression (\ref{Psi4_ExpressionDef}) of $\Psi_4$. However, when considering the term $\Psi^{\text{\tiny{RP}} \, \textit{\tiny{(P)}}}_4$ alone, we shall encounter some problems due to the pathological behavior of $Z_{\lambda_n(\omega) -1/2} (\cos \varphi)$ for $\varphi \to \pi$. In fact, both the Regge pole approximation $\Psi^{\text{\tiny{RP}} \, \textit{\tiny{(P)}}}_4$ and the background integral contribution $\Psi^{\text{\tiny{B}} \, \textit{\tiny{(P)}}}_4$ are divergent in the limit $\varphi \to \pi$ but it is worth pointing out that their sum (\ref{CAM_Psi4_ExpressionDef_Ptot}) does not present any pathology.

\subsection{CAM representation and Regge pole approximation of the Weyl scalar $\Psi_4$ based on the Sommerfeld-Watson transform}
\label{SecIIId}

By means of the Sommerfeld-Watson transformation \cite{Watson18,Sommerfeld49,Newton:1982qc} which permits us to write
\begin{equation}\label{SWT_gen}
\sum_{\ell=0}^{+\infty} (-1)^\ell F(\ell)= \frac{i}{2} \int_{\cal C} d\lambda \, \frac{F(\lambda -1/2)}{\cos (\pi \lambda)},
\end{equation}
we replace in Eq.~(\ref{Psi4_ExpressionDef_ter}) the discrete sum over the ordinary angular momentum $\ell$ by a contour integral in the
complex $\lambda$ plane (i.e., in the complex $\ell$ plane with $\lambda = \ell +1/2$). We obtain
\begin{eqnarray}\label{SW_Psi4_ExpressionDef}
& & r\, \Psi_4 (t,r,\theta=\pi/2,\varphi)= \frac{1}{\sqrt{2\pi}} \int_{-\infty}^{+\infty} d\omega \, e^{-i\omega[ t-r_\ast(r)]} \nonumber \\
& & \qquad \times \left[ \frac{i}{2} \int_{\cal C} d\lambda \, \frac{\lambda}{2\pi \cos (\pi \lambda)} \phantom{\frac{ K[\lambda-1/2,\omega]}{A^{(-)}_{\lambda-1/2} (\omega)}}  \right. \nonumber \\
& & \qquad\qquad  \left. \times \frac{i \omega K[\lambda-1/2,\omega]}{4 A^{(-)}_{\lambda-1/2} (\omega)}  \, Z_{\lambda -1/2} (-\cos \varphi) \right].
\end{eqnarray}
In Eqs.~(\ref{SWT_gen}) and (\ref{SW_Psi4_ExpressionDef}), the integration contour encircles counterclockwise the positive real semiaxis of the complex $\lambda$ plane, i.e., we take ${\cal C}=]+\infty +i\epsilon,+i\epsilon] \cup
[+i\epsilon,-i\epsilon] \cup [-i\epsilon, +\infty -i\epsilon[$
with $\epsilon \to 0_+$. We can recover
(\ref{Psi4_ExpressionDef_ter}) from (\ref{SW_Psi4_ExpressionDef}) by using
Cauchy's residue theorem and by noting that the poles of the integrand in
(\ref{SW_Psi4_ExpressionDef}) that are enclosed into ${\cal C}$ are the zeros of $\cos (\pi \lambda)$, i.e., the semi-integers $\lambda = \ell + 1/2$ with $\ell \in \mathbb{N}$.

The contour ${\cal C}$ in Eq.~(\ref{SW_Psi4_ExpressionDef}) is now open out to become a contour running on the imaginary axis of the complex $\lambda$ plane (for more details, see, e.g., Ref.~\cite{Newton:1982qc}). This permits us to collect, by using Cauchy's residue theorem, the contributions from the Regge poles $\lambda_n(\omega)$ with $n=1, 2, 3, \dots  $ that are crossed over. Here, it should be recalled that these poles of the ${\cal S}$-matrix lie in the first quadrant of the CAM plane for $\omega >0$ and that they migrate in the fourth one for $\omega <0$ (see also Sec.~\ref{SecIIIb}). By again assuming that the contributions coming from the quarter circles at infinity vanish, we obtain
\begin{widetext}
\begin{equation}\label{CAM_Psi4_ExpressionDef_SWtot}
\Psi_4 (t,r,\theta=\pi/2,\varphi)= \Psi^{\text{\tiny{B}} \, \textit{\tiny{(SW)}}}_4 (t,r,\theta=\pi/2,\varphi) + \Psi^{\text{\tiny{RP}} \, \textit{\tiny{(SW)}}}_4 (t,r,\theta=\pi/2,\varphi)
\end{equation}
where
\begin{subequations}\label{CAM_Psi4_ExpressionDef_SW}
\begin{equation}\label{CAM_Psi4_ExpressionDef_SW_Background}
r\, \Psi^{\text{\tiny{B}} \, \textit{\tiny{(SW)}}}_4 (t,r,\theta=\pi/2,\varphi)= \frac{1}{\sqrt{2\pi}} \int_{-\infty}^{+\infty} d\omega \, e^{-i\omega[ t-r_\ast(r)]} \left[\frac{1}{16\pi} \int_{-i\infty}^{+i\infty} d\lambda \, \frac{\lambda}{\cos (\pi \lambda)} \, \frac{\omega K[\lambda-1/2,\omega]}{ A^{(-)}_{\lambda-1/2} (\omega)} \,  Z_{\lambda -1/2} (-\cos \varphi) \right]
\end{equation}
is a background integral contribution and where
\begin{eqnarray}\label{CAM_Psi4_ExpressionDef_SW_RP}
& & r\, \Psi^{\text{\tiny{RP}} \, \textit{\tiny{(SW)}}}_4 (t,r,\theta=\pi/2,\varphi)= \frac{1}{\sqrt{2\pi}} \int_{-\infty}^{+\infty} d\omega \, e^{-i\omega[ t-r_\ast(r)]}   \nonumber \\
& & \qquad\qquad\qquad\qquad\qquad   \times \left[ \sum_{n=1}^{+\infty}  \frac{\lambda_n(\omega) \beta_n(\omega)}{4 \,\cos[\pi \lambda_n(\omega)]}  \, \frac{i \omega^2 K[\lambda_n(\omega)-1/2,\omega]}{ A^{(+)}_{\lambda_n(\omega)-1/2} (\omega)} \,  Z_{\lambda_n(\omega) -1/2} (-\cos \varphi) \right]
\end{eqnarray}
\end{subequations}
\end{widetext}
is the Fourier transform of a sum over the Regge poles. We can again check that $\Psi_4$ is a real-valued function by now considering this last expression. Indeed, due to the relations (\ref{prop_ext_Z_b}) and (\ref{Sym_om_CAM_d}), the term into the squared brackets in the r.h.s.~of (\ref{CAM_Psi4_ExpressionDef_SW_Background}) satisfies the Hermitian symmetry property. Such a property is also satisfied by the term into the squared brackets in the r.h.s.~of (\ref{CAM_Psi4_ExpressionDef_SW_RP}) as a consequence of the relations (\ref{prop_ext_Z_b}), (\ref{Sym_om_CAM_b}), (\ref{Sym_om_CAM_c}), (\ref{PR_chgt_om}) and (\ref{EF_PR_chgt_om}).

Of course, Eq.~(\ref{CAM_Psi4_ExpressionDef_SWtot}) provides an exact expression for the Weyl scalar $\Psi_4$, equivalent to the initial expression (\ref{Psi4_ExpressionDef}) and to the expression (\ref{CAM_Psi4_ExpressionDef_P}) obtained from the Poisson summation formula. From this CAM representation of $\Psi_4$, we can extract the contribution denoted by $\Psi^{\text{\tiny{RP}} \, \textit{\tiny{(SW)}}}_4$ and given by (\ref{CAM_Psi4_ExpressionDef_SW_RP}) which, as a sum over the Regge poles, is only an approximation of $\Psi_4$. In Sec.~\ref{SecIV}, we shall compare it with the exact expression (\ref{Psi4_ExpressionDef}) of $\Psi_4$ and with the Regge pole approximation $\Psi^{\text{\tiny{RP}} \, \textit{\tiny{(P)}}}_4$ obtained in Sec.~\ref{SecIIIc}. However, when considering the term $\Psi^{\text{\tiny{RP}} \, \textit{\tiny{(SW)}}}_4$ alone, we shall encounter some problems due to the pathological behavior of  $Z_{\lambda_n(\omega) -1/2} (-\cos \varphi)$ for $\varphi \to 0$. In fact, both the Regge pole approximation $\Psi^{\text{\tiny{RP}} \, \textit{\tiny{(SW)}}}_4$ and the background integral contribution $\Psi^{\text{\tiny{B}} \, \textit{\tiny{(SW)}}}_4$ are divergent in the limit $\varphi \to 0$ but it is worth pointing out that their sum (\ref{CAM_Psi4_ExpressionDef_SWtot}) does not present any pathology.

\section{Comparison of the Weyl scalar $\Psi_4$ with its Regge pole approximations}
\label{SecIV}

In this section, we shall compare numerically the multipolar waveform $\Psi_4$ given by (\ref{Psi4_ExpressionDef}) and constructed by summing over a large number of partial modes (this is particularly necessary for the radially infalling relativist particle) as well as the associated quasinormal ringdown $\Psi^\text{\tiny{QNM}}_4$ given by (\ref{response_QNM}) with the Regge pole approximations $\Psi^{\text{\tiny{RP}} \, \textit{\tiny{(P)}}}_4$ and $\Psi^{\text{\tiny{RP}} \, \textit{\tiny{(SW)}}}_4$ respectively given by (\ref{CAM_Psi4_ExpressionDef_P_RP}) and (\ref{CAM_Psi4_ExpressionDef_SW_RP}) and constructed by considering only one or a small number of Regge poles. This will allow us to clearly highlight the benefits of working with the Regge pole approximations of $\Psi_4$.

\subsection{Numerical methods}
\label{SecIVa}

To construct numerically the Weyl scalar $\Psi_4$ as well as its QNM and Regge pole approximations:
\begin{enumerate}[label=(\arabic*)]

\item  We have to solve the problem (\ref{H_ZM_equation})-(\ref{bc_in}) permitting us to obtain the function $\phi_{\omega, \ell}^{\mathrm {in}}(r)$ and the coefficients $A^{(-)}_\ell (\omega)$ and  $A^{(+)}_\ell (\omega)$. This must be achieved (i) for $\ell \in \mathbb{N}$ and $\omega \in \mathbb{R}$ as well as (ii) for $\ell \in \mathbb{N}$ and $\omega \in \mathbb{C}$ [or, more precisely, for the quasinormal frequencies $\omega=\omega_{\ell n}$] and (iii) for $\ell=\lambda-1/2 \in \mathbb{C}$ [or, more precisely, for the Regge poles $\lambda=\lambda_n(\omega)$] and $\omega \in \mathbb{R}$.

\item We have to determine the quasinormal frequencies $\omega_{\ell n}$ and the Regge poles $\lambda_n(\omega)$, i.e., the solutions of (\ref{freqQN_def_Am}) and (\ref{PR_def_Am}), and to obtain the corresponding excitation factors (\ref{excitation_factor_QNM}) and (\ref{excitation_factor_RP}). Let us recall that this point has been already discussed in Sec.~\ref{SecIIIb}.

\item We have to construct the term $K[\ell,\omega]$ defined by (\ref{Partial_Response_b}) (i) for $\ell \in \mathbb{N}$ and $\omega \in \mathbb{R}$ as well as (ii) for $\ell \in \mathbb{N}$ and $\omega =\omega_{\ell n}$ and (iii) for $\ell=\lambda_n(\omega)-1/2$ and $\omega \in \mathbb{R}$.

\end{enumerate}
All these numerical results can be obtained by using, {\it mutatis mutandis}, the methods that have permitted us to describe, in Refs.~\cite{Folacci:2018vtf} and \cite{Folacci:2018cic}, the electromagnetic field and the gravitational waves generated by a particle plunging from the innermost stable circular orbit (ISCO) into a Schwarzschild BH (see, e.g., Sec.~IVA of Ref.~\cite{Folacci:2018cic}). It should be noted, however, that, in these previous works, it has been necessary to regularize the multipolar waveforms due to divergences occurring near the ISCO. We do not encounter such a problem for the particle falling radially from infinity into a Schwarzschild BH. It should also be recalled that it is necessary to select a starting time $t_\mathrm{start}$ for the ringdown. By taking $t_\mathrm{start}=t_p(3M)$, i.e., the moment the particle crosses the photon sphere, we have obtained physically relevant results. Finally, it should be noted that we have performed all the numerical calculations by using {\it Mathematica} \cite{Mathematica} and by taking $t_0=0$ in Eqs.~(\ref{trajectory_Rad}) and (\ref{trajectory_RadRel}).

\begin{figure*}
\centering
 \includegraphics[scale=0.50]{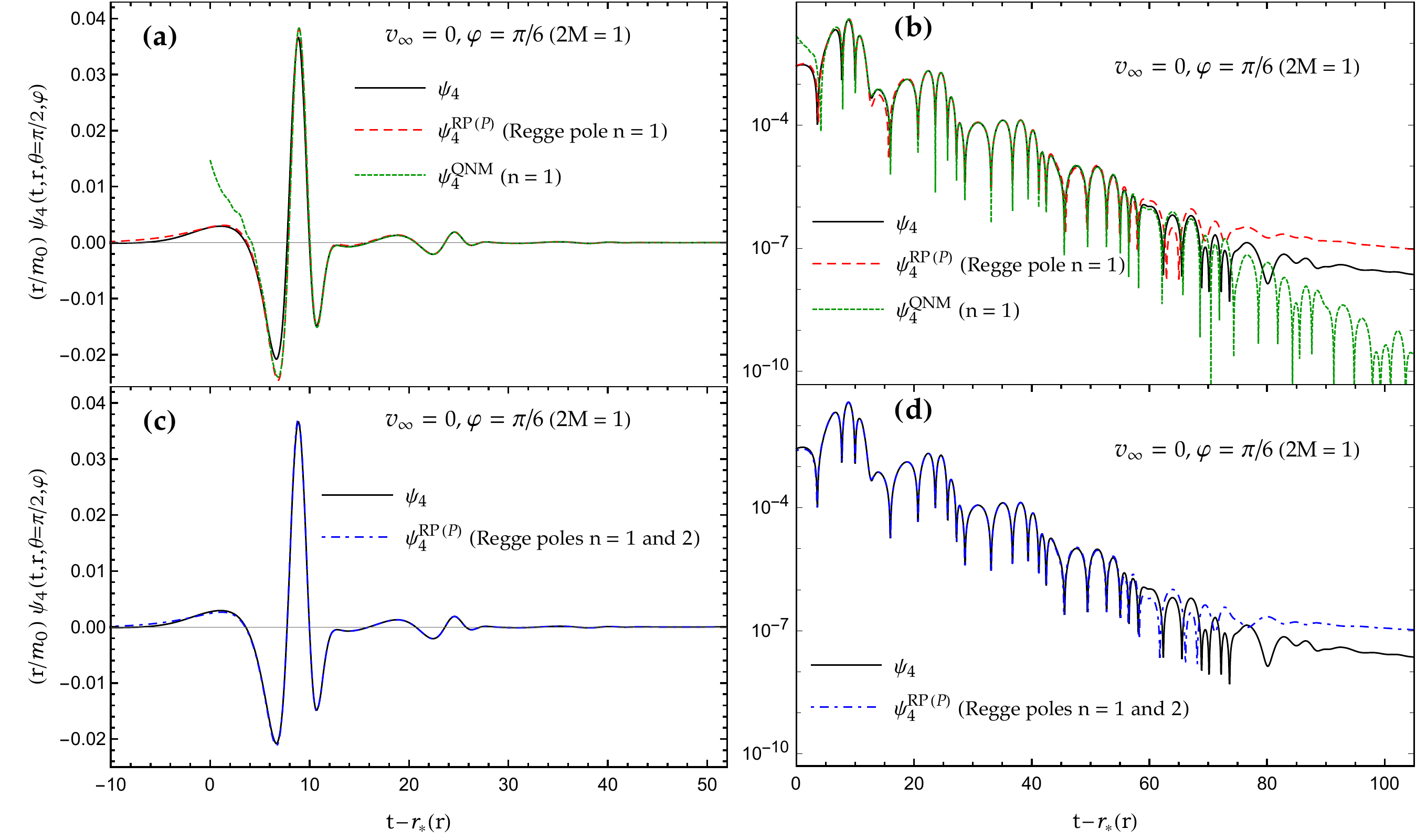}
\caption{\label{P_Exact_QNM_CAM_pis6} The Weyl scalar $\Psi_4$ and its Regge pole approximation $\Psi^{\text{\tiny{RP}} \, \textit{\tiny{(P)}}}_4$ for $v_\infty =0$ ($\gamma=1$) and $\varphi=\pi/6$. (a) The Regge pole approximation constructed from only one Regge pole is in very good agreement with the Weyl scalar $\Psi_4$ constructed by summing over the first eight partial waves. The associated quasinormal response $\Psi^\text{\tiny{QNM}}_4$  obtained by summing over the $(\ell, n)$ QNMs with $n=1$ and $\ell=2, \dots, 10$ is also displayed. At intermediate timescales, it matches very well the Regge pole approximation. (b) Semilog graph corresponding to (a) and showing that the Regge pole approximation describes very well the ringdown and roughly approximates the waveform tail. (c) Taking into account an additional Regge pole slightly improves the Regge pole approximation which is now in perfect agreement with the Weyl scalar $\Psi_4$. (d) Semilog graph corresponding to (c) and showing that the additional Regge pole does not improve the description of the waveform tail.}
\end{figure*}

\begin{figure*}
\centering
 \includegraphics[scale=0.50]{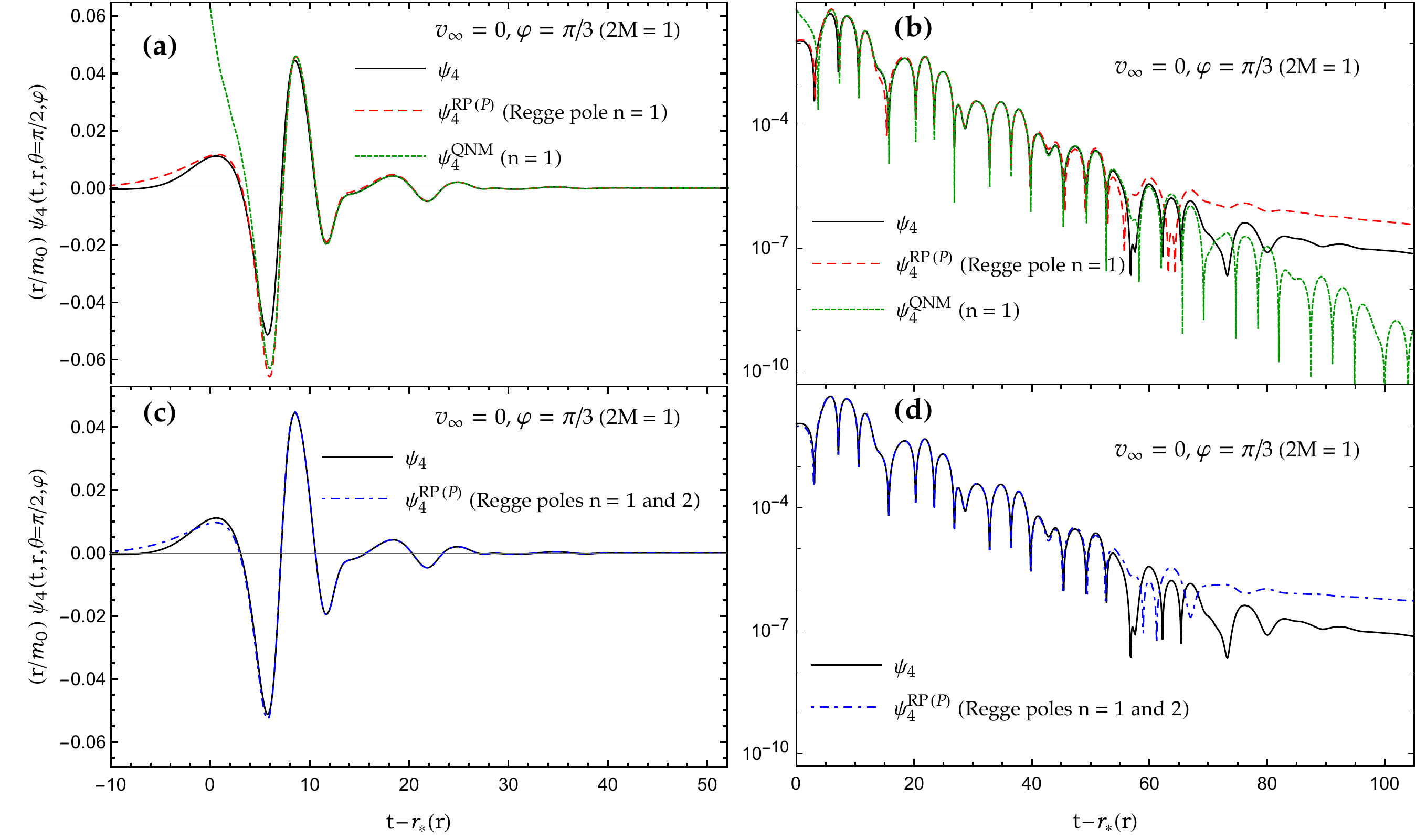}
\caption{\label{P_Exact_QNM_CAM_pis3} The Weyl scalar $\Psi_4$ and its Regge pole approximation $\Psi^{\text{\tiny{RP}} \, \textit{\tiny{(P)}}}_4$ for $v_\infty =0$ ($\gamma=1$) and $\varphi=\pi/3$. (a) The Regge pole approximation constructed from only one Regge pole is in good agreement with the Weyl scalar $\Psi_4$ constructed by summing over the first eight partial waves. The associated quasinormal response $\Psi^\text{\tiny{QNM}}_4$  obtained by summing over the $(\ell, n)$ QNMs with $n=1$ and $\ell=2, \dots, 10$ is also displayed. At intermediate timescales, it matches very well the Regge pole approximation. (b) Semilog graph corresponding to (a) and showing that the Regge pole approximation describes very well the ringdown and roughly approximates the waveform tail. (c) Taking into account an additional Regge pole slightly improves the Regge pole approximation which is now in very good agreement with the Weyl scalar $\Psi_4$. (d) Semilog graph corresponding to (c) and showing that the additional Regge pole does not improve the description of the waveform tail.}
\end{figure*}

\begin{figure*}
\centering
 \includegraphics[scale=0.50]{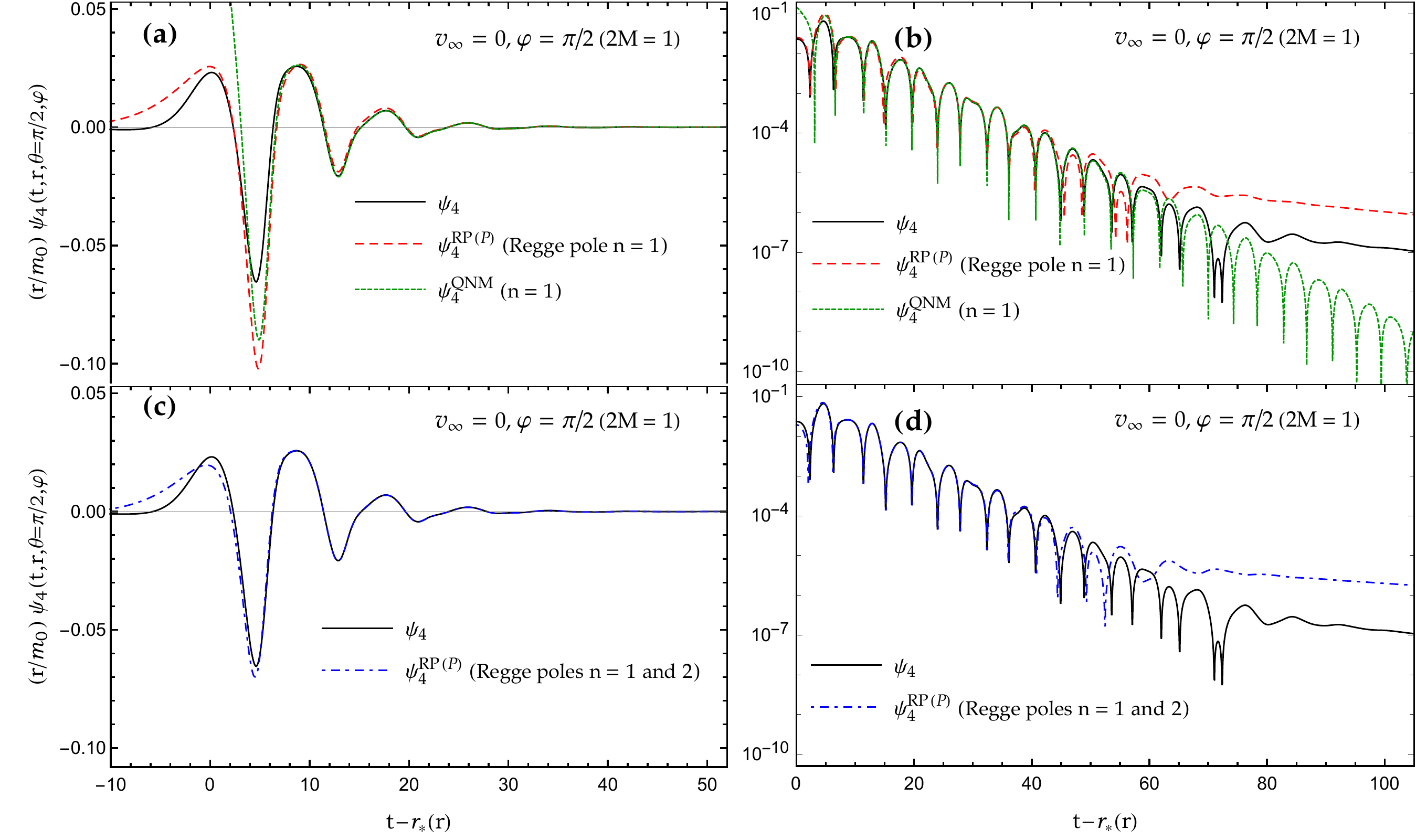}
\caption{\label{P_Exact_QNM_CAM_pis2} The Weyl scalar $\Psi_4$ and its Regge pole approximation $\Psi^{\text{\tiny{RP}} \, \textit{\tiny{(P)}}}_4$ for $v_\infty =0$ ($\gamma=1$) and $\varphi=\pi/2$. (a) The Regge pole approximation constructed from only one Regge pole is in rather good agreement with the Weyl scalar $\Psi_4$ constructed by summing over the first eight partial waves. The associated quasinormal response $\Psi^\text{\tiny{QNM}}_4$ obtained by summing over the $(\ell, n)$ QNMs with $n=1$ and $\ell=2, \dots, 10$ is also displayed. At intermediate timescales, it matches very well the Regge pole approximation. (b) Semilog graph corresponding to (a) and showing that the Regge pole approximation describes very well a large part of the ringdown and roughly approximates the waveform tail. (c) Taking into account an additional Regge pole improves the Regge pole approximation which is now in very good agreement with the Weyl scalar $\Psi_4$. (d) Semilog graph corresponding to (c) and showing that the additional Regge pole does not improve the description of the waveform tail.}
\end{figure*}

\begin{figure*}
\centering
 \includegraphics[scale=0.50]{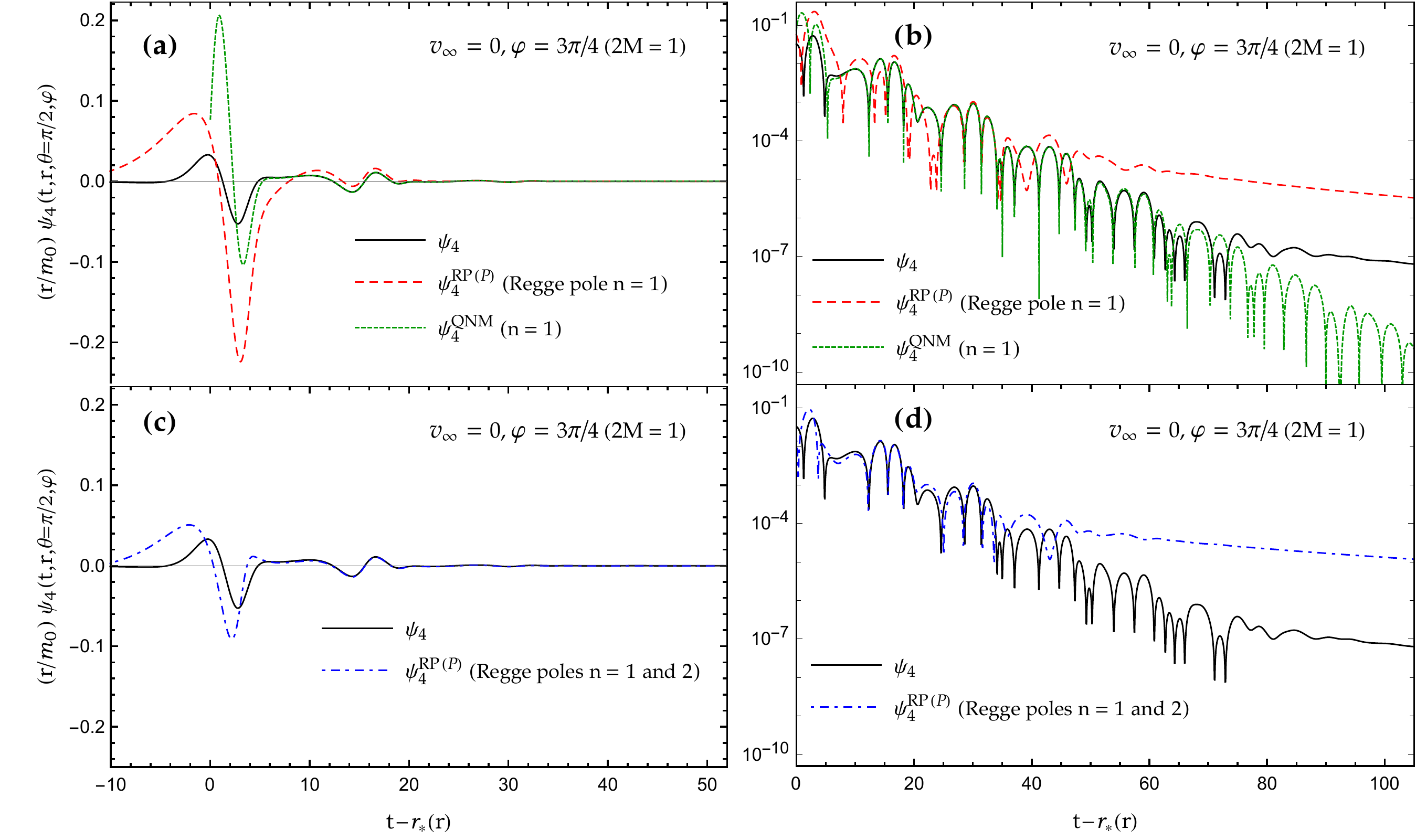}
\caption{\label{P_Exact_QNM_CAM_3pis4} The Weyl scalar $\Psi_4$ and its Regge pole approximation $\Psi^{\text{\tiny{RP}} \, \textit{\tiny{(P)}}}_4$ for $v_\infty =0$ ($\gamma = 1$) and $\varphi=3\pi/4$. (a) The Regge pole approximation constructed from only one Regge pole does not match the Weyl scalar $\Psi_4$ constructed by summing over the first eight partial waves. The associated quasinormal response $\Psi^\text{\tiny{QNM}}_4$  obtained by summing over the $(\ell, n)$ QNMs with $n=1$ and $\ell=2, \dots, 10$ is also displayed. The discrepancy with the Regge pole approximation is obvious. (b) Semilog graph corresponding to (a) and showing that the Regge pole approximation describes correctly a small part of the ringdown. (c) Taking into account an additional Regge pole slightly improves the Regge pole approximation which is now in rough agreement with the Weyl scalar $\Psi_4$. (d) Semilog graph corresponding to (c) and showing that the Regge pole approximation correctly describes a rather large part of the ringdown.}
\end{figure*}

 \begin{figure*}
\centering
 \includegraphics[scale=0.50]{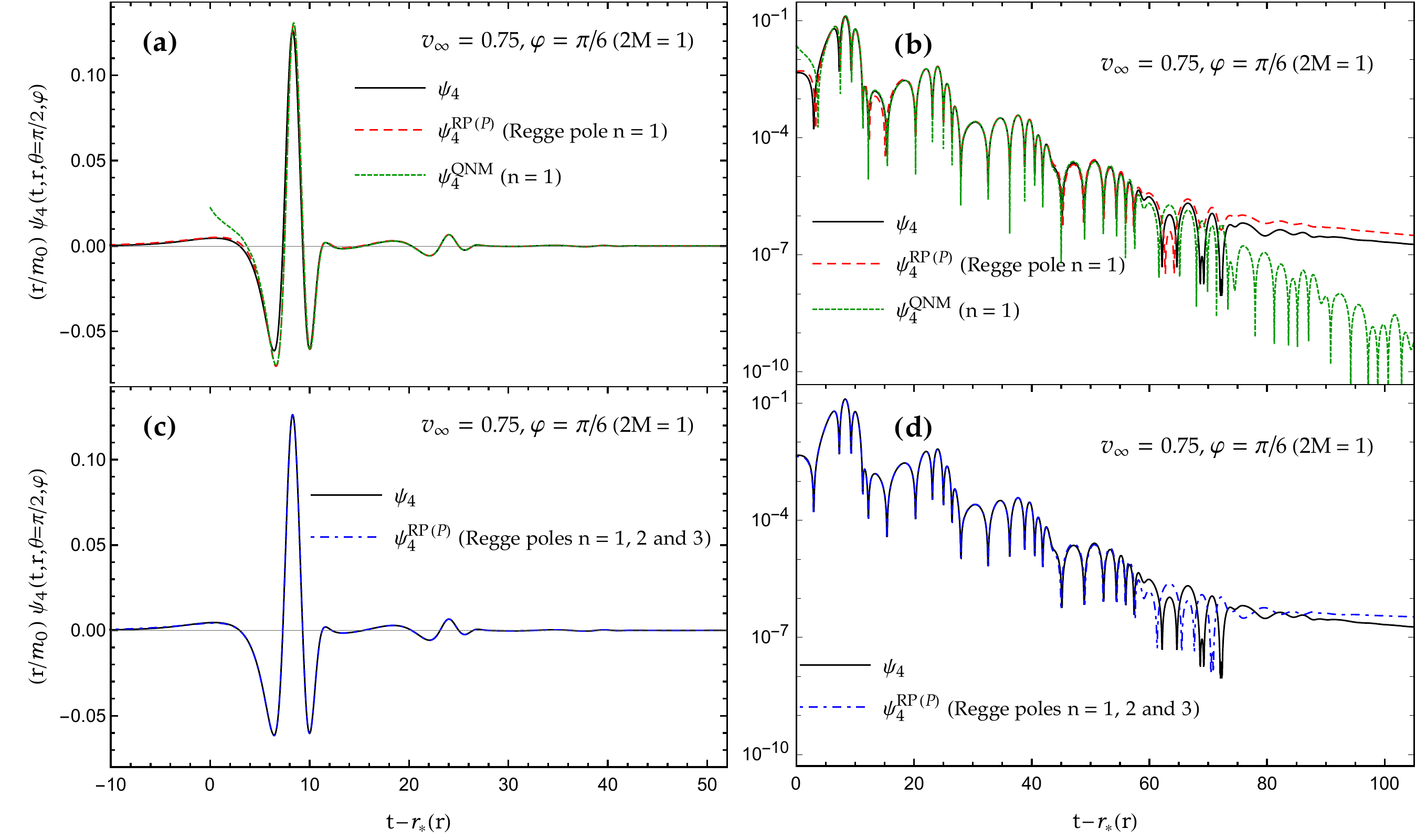}
\caption{\label{P_Exact_QNM_CAM_pis6_R} The Weyl scalar $\Psi_4$ and its Regge pole approximation $\Psi^{\text{\tiny{RP}} \, \textit{\tiny{(P)}}}_4$ for $v_\infty =0.75$ ($\gamma \approx 1.51$) and $\varphi=\pi/6$. (a) The Regge pole approximation constructed from only one Regge pole is in very good agreement with the Weyl scalar $\Psi_4$ constructed by summing over the first thirteen partial waves. The associated quasinormal response $\Psi^\text{\tiny{QNM}}_4$  obtained by summing over the $(\ell, n)$ QNMs with $n=1$ and $\ell=2, \dots, 15$ is also displayed. At intermediate timescales, it matches very well the Regge pole approximation. (b) Semilog graph corresponding to (a) and showing that the Regge pole approximation describes very well the ringdown and correctly the waveform tail. (c) Taking into account two additional Regge poles slightly improves the Regge pole approximation which is now in perfect agreement with the Weyl scalar $\Psi_4$. (d) Semilog graph corresponding to (c) and showing that the additional Regge poles slightly improves the description of the waveform tail.}
\end{figure*}

 \begin{figure*}
\centering
 \includegraphics[scale=0.50]{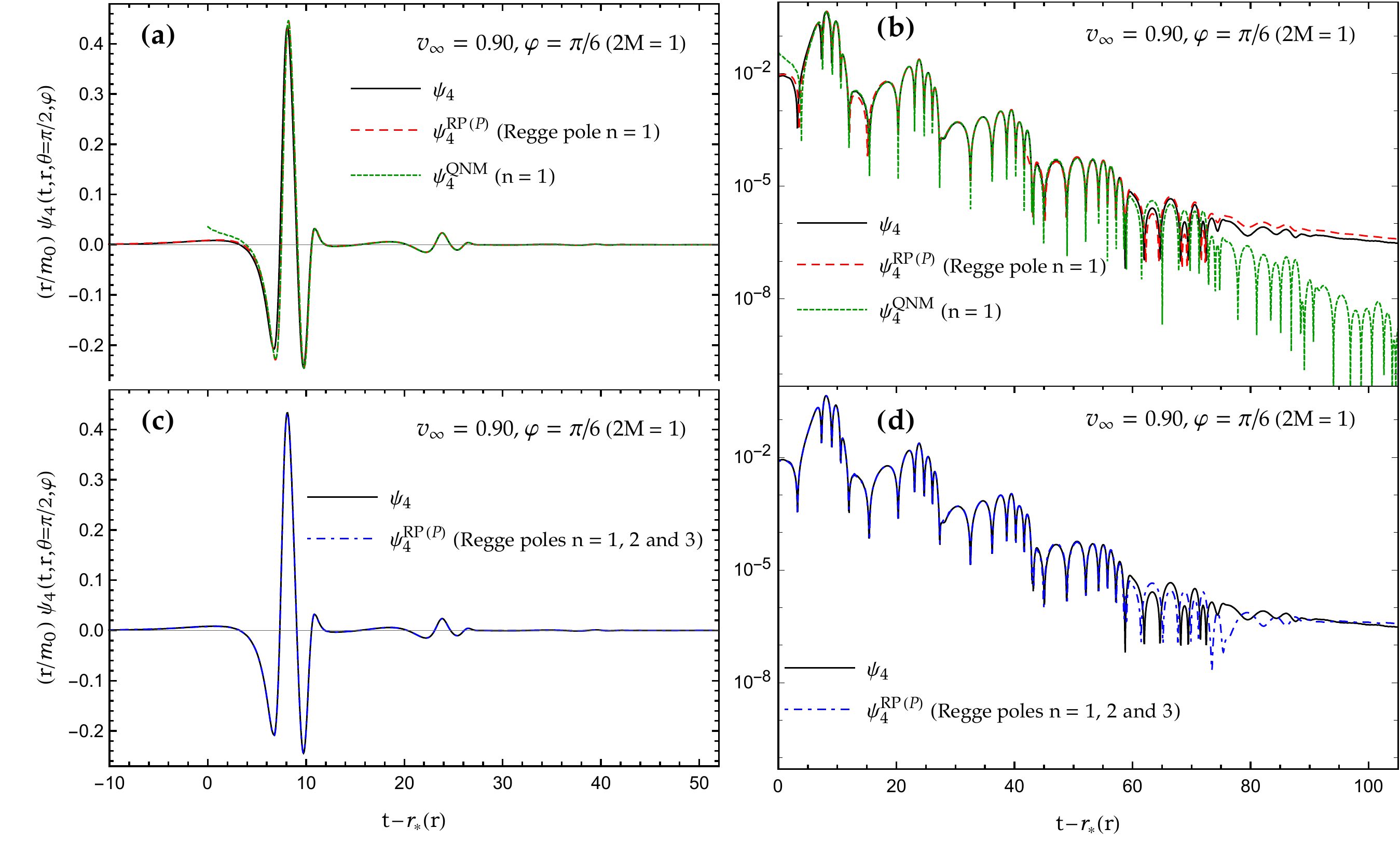}
\caption{\label{P_Exact_QNM_CAM_pis6_UR} The Weyl scalar $\Psi_4$ and its Regge pole approximation $\Psi^{\text{\tiny{RP}} \, \textit{\tiny{(P)}}}_4$ for $v_\infty =0.90$ ($\gamma\approx 2.29$) and $\varphi=\pi/6$. (a) The Regge pole approximation constructed from only one Regge pole is in very good agreement with the Weyl scalar $\Psi_4$ constructed by summing over the first eighteen partial waves. The associated quasinormal response $\Psi^\text{\tiny{QNM}}_4$  obtained by summing over the $(\ell, n)$ QNMs with $n=1$ and $\ell=2, \dots, 20$ is also displayed. At intermediate timescales, it matches very well the Regge pole approximation. (b) Semilog graph corresponding to (a) and showing that the Regge pole approximation describes very well the ringdown and the waveform tail. (c) Taking into account two additional Regge poles slightly improves the Regge pole approximation which is now in perfect agreement with the Weyl scalar $\Psi_4$. (d) Semilog graph corresponding to (c) and showing that the whole signal is impressively described by the Regge pole approximation.}
\end{figure*}

 \begin{figure*}
\centering
 \includegraphics[scale=0.50]{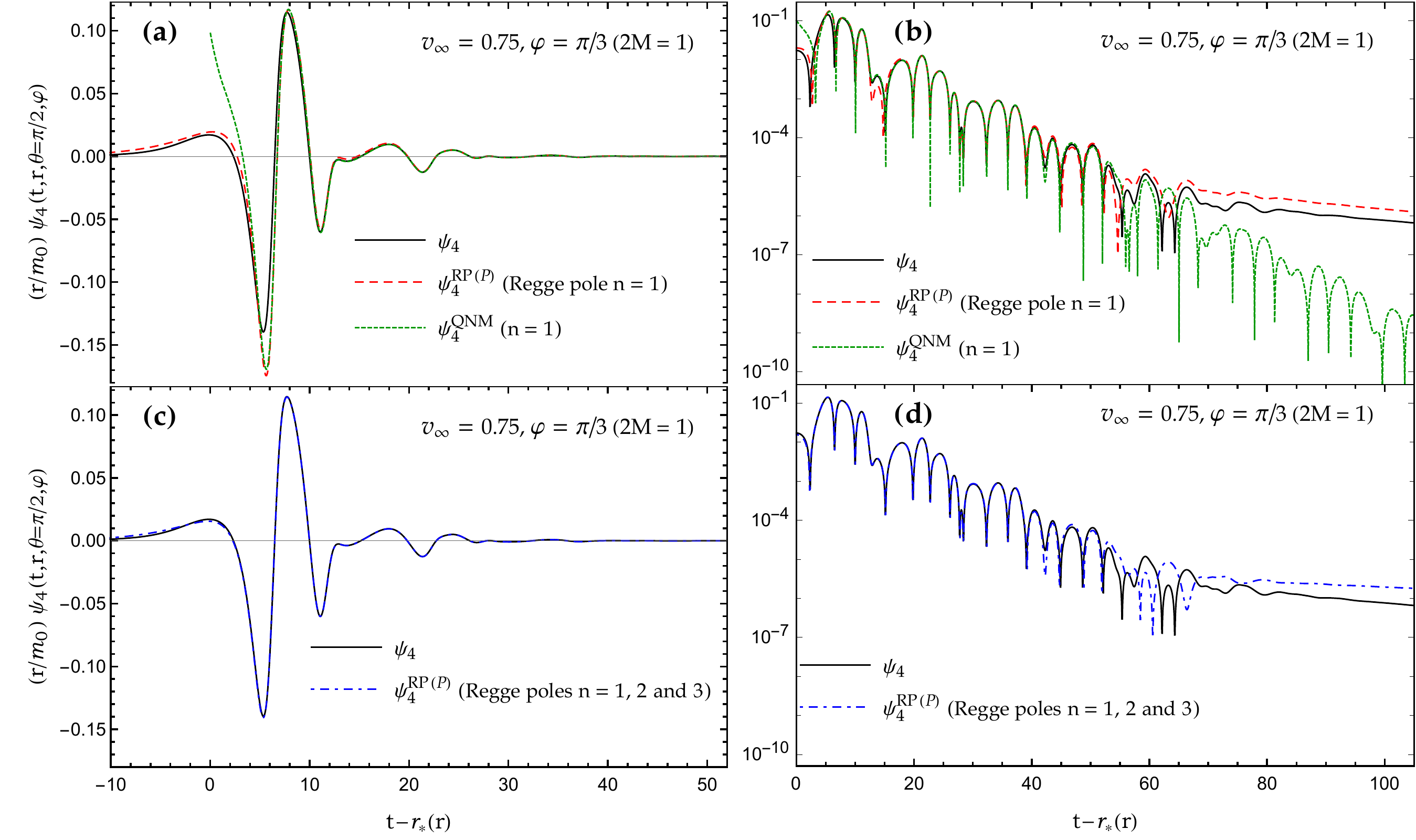}
\caption{\label{P_Exact_QNM_CAM_pis3_R} The Weyl scalar $\Psi_4$ and its Regge pole approximation $\Psi^{\text{\tiny{RP}} \, \textit{\tiny{(P)}}}_4$ for $v_\infty =0.75$ ($\gamma \approx 1.51$) and $\varphi=\pi/3$. (a) The Regge pole approximation constructed from only one Regge pole is in good agreement with the Weyl scalar $\Psi_4$ constructed by summing over the first thirteen partial waves. The associated quasinormal response $\Psi^\text{\tiny{QNM}}_4$  obtained by summing over the $(\ell, n)$ QNMs with $n=1$ and $\ell=2, \dots, 15$ is also displayed. At intermediate timescales, it matches very well the Regge pole approximation. (b) Semilog graph corresponding to (a) and showing that the Regge pole approximation describes very well the ringdown and correctly the waveform tail. (c) Taking into account two additional Regge poles slightly improves the Regge pole approximation which is now in perfect agreement with the Weyl scalar $\Psi_4$. (d) Semilog graph corresponding to (c) and showing that the additional Regge poles slightly improves the description of the waveform.}
\end{figure*}

\begin{figure*}
\centering
 \includegraphics[scale=0.50]{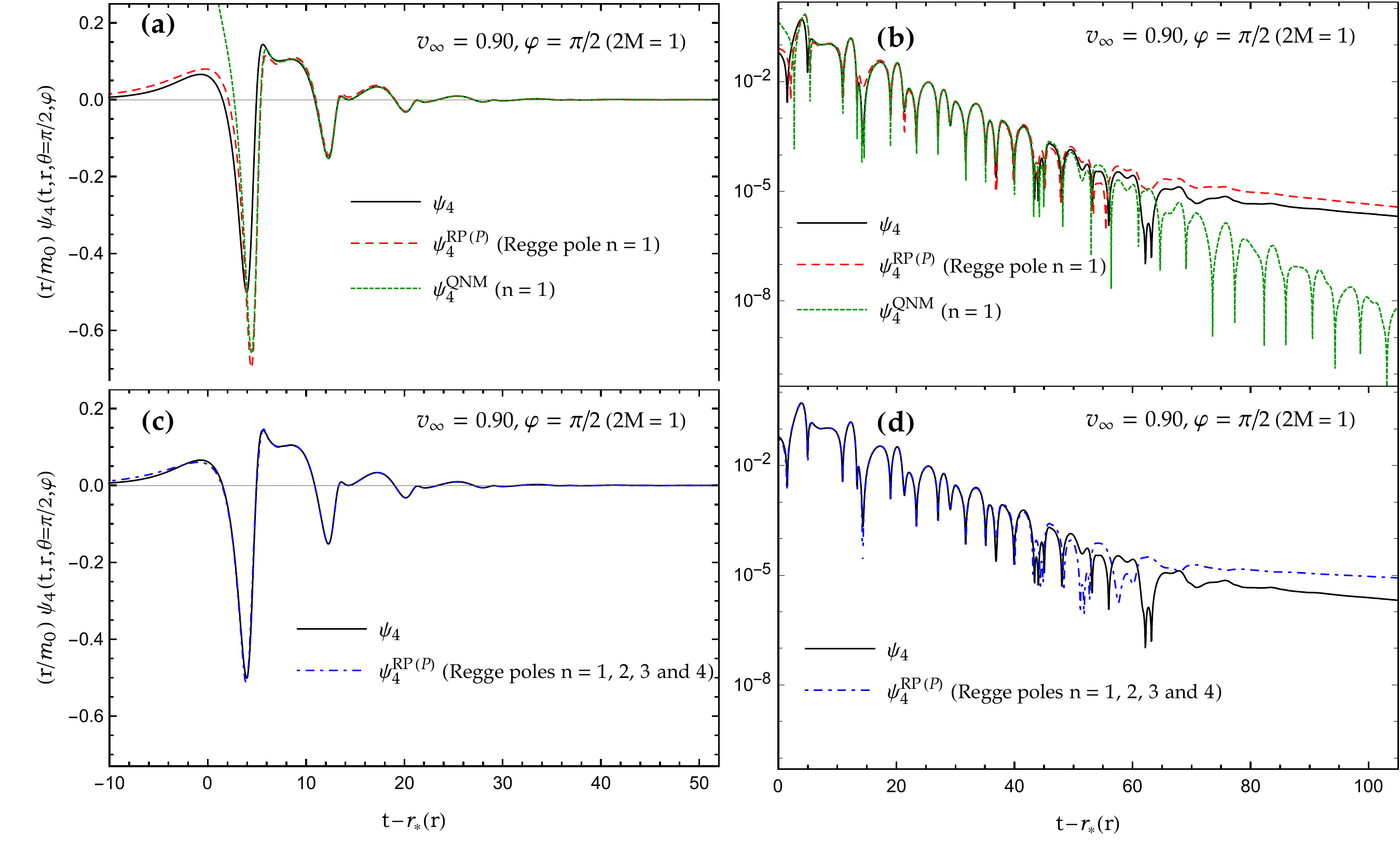}
\caption{\label{P_Exact_QNM_CAM_pis2_R_avec_4} The Weyl scalar $\Psi_4$ and its Regge pole approximation $\Psi^{\text{\tiny{RP}} \, \textit{\tiny{(P)}}}_4$ for $v_\infty =0.90$ ($\gamma\approx 2.29$) and $\varphi=\pi/2$. (a) The Regge pole approximation constructed from only one Regge pole is in rather good agreement with the Weyl scalar $\Psi_4$ constructed by summing over the first eighteen partial waves. The associated quasinormal response $\Psi^\text{\tiny{QNM}}_4$  obtained by summing over the $(\ell, n)$ QNMs with $n=1$ and $\ell=2, \dots, 20$ is also displayed. At intermediate timescales, it matches very well the Regge pole approximation. (b) Semilog graph corresponding to (a) and showing that the Regge pole approximation describes very well the ringdown and correctly the waveform tail. (c) Taking into account three additional Regge poles improves the Regge pole approximation which is now in very good agreement with the Weyl scalar $\Psi_4$. (d) Semilog graph corresponding to (c) and showing that a large part of the signal is very well described by the Regge pole approximation.}
\end{figure*}

\begin{figure*}
\centering
 \includegraphics[scale=0.50]{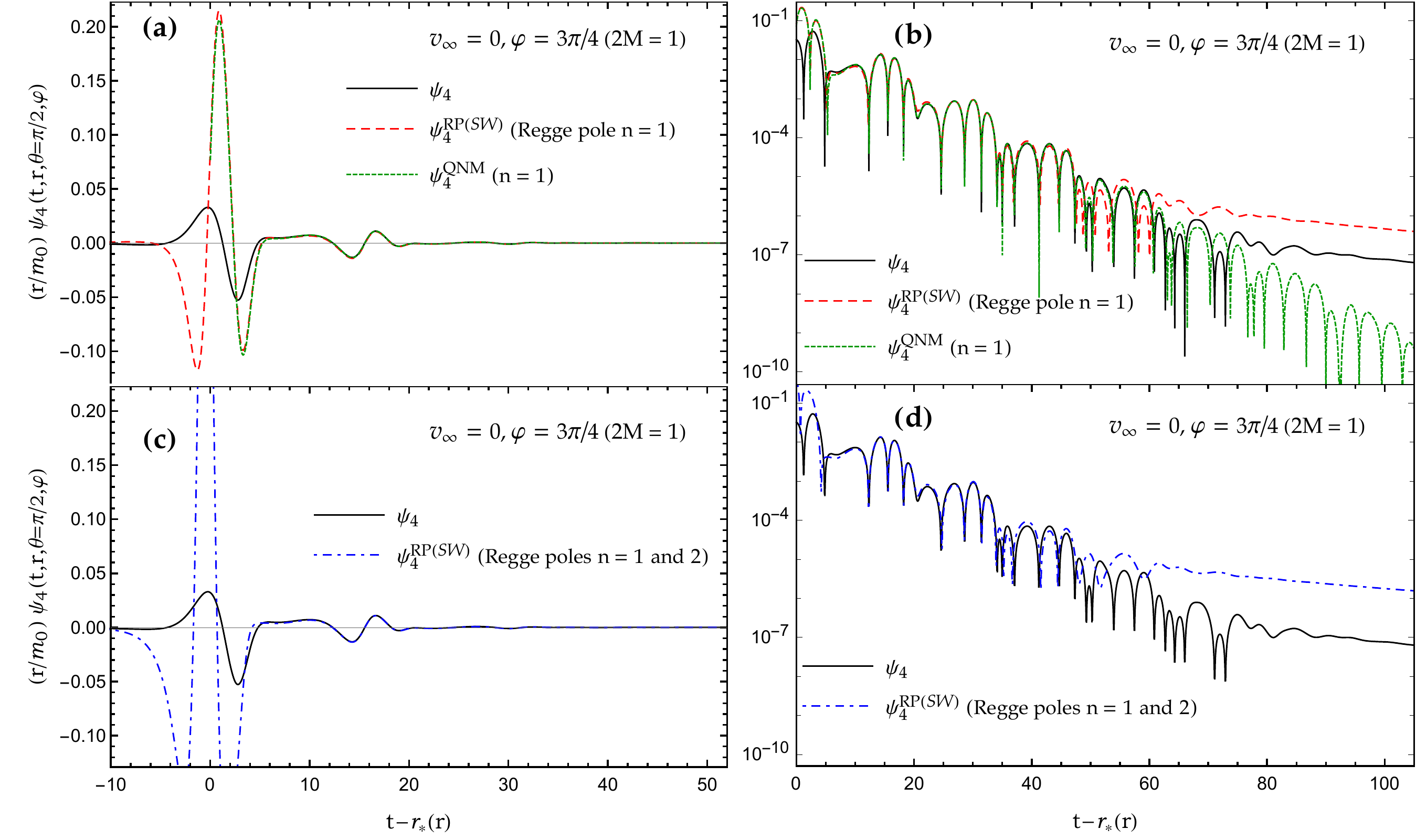}
\caption{\label{SW_Exact_QNM_CAM_3pis4} The Weyl scalar $\Psi_4$ and its Regge pole approximation $\Psi^{\text{\tiny{RP}} \, \textit{\tiny{(SW)}}}_4$ for $v_\infty =0$ ($\gamma\approx 1$) and $\varphi=3\pi/4$. (a) and (b) The pre-ringdown phase of the Weyl scalar $\Psi_4$ constructed by summing over the first eight partial waves is not described by the Regge pole approximation constructed from only one Regge pole. However, this approximation matches a large part of the ringdown and roughly approximates the waveform tail. The quasinormal response $\Psi^\text{\tiny{QNM}}_4$  obtained by summing over the $(\ell, n)$ QNMs with $n=1$ and $\ell=2, \dots, 10$ is also displayed. At intermediate timescales, it matches very well the Regge pole approximation. (c) and (d) Taking into account an additional Regge pole does not improve the Regge pole approximation.}
\end{figure*}

\begin{figure*}
\centering
 \includegraphics[scale=0.50]{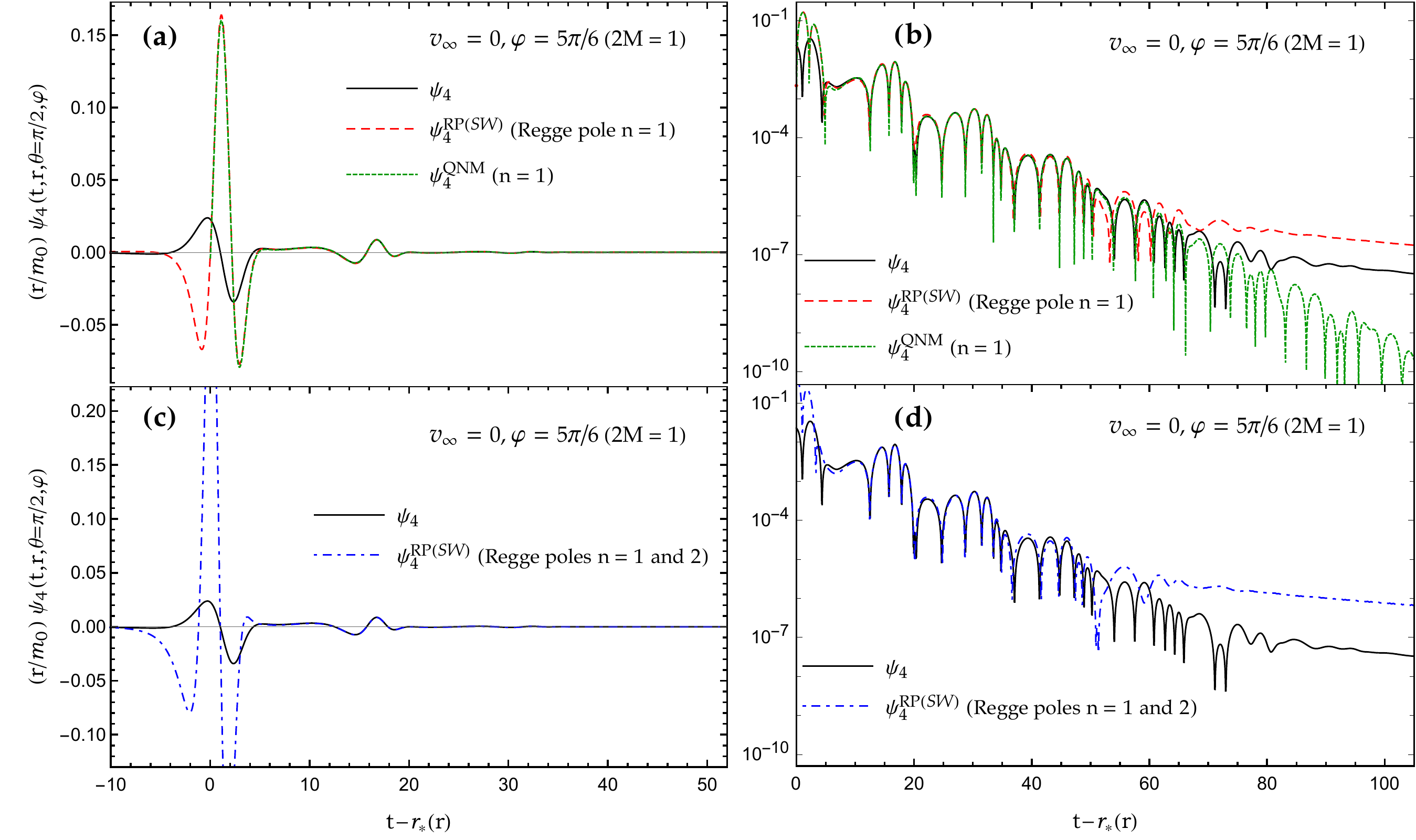}
\caption{\label{SW_Exact_QNM_CAM_5pis6} The Weyl scalar $\Psi_4$ and its Regge pole approximation $\Psi^{\text{\tiny{RP}} \, \textit{\tiny{(SW)}}}_4$ for $v_\infty =0$ ($\gamma=1$) and $\varphi=5\pi/6$. (a) and (b) The pre-ringdown phase of the Weyl scalar $\Psi_4$ constructed by summing over the first eight partial waves is not described by the Regge pole approximation constructed from only one Regge pole. By contrast, this approximation matches a large part of the ringdown and roughly approximates the waveform tail. The quasinormal response $\Psi^\text{\tiny{QNM}}_4$ obtained by summing over the $(\ell, n)$ QNMs with $n=1$ and $\ell=2, \dots, 10$ is also displayed. At intermediate timescales, it matches very well the Regge pole approximation. (c) and (d) Taking into account an additional Regge pole does not improve the Regge pole approximation.}
\end{figure*}

 \begin{figure*}
\centering
 \includegraphics[scale=0.50]{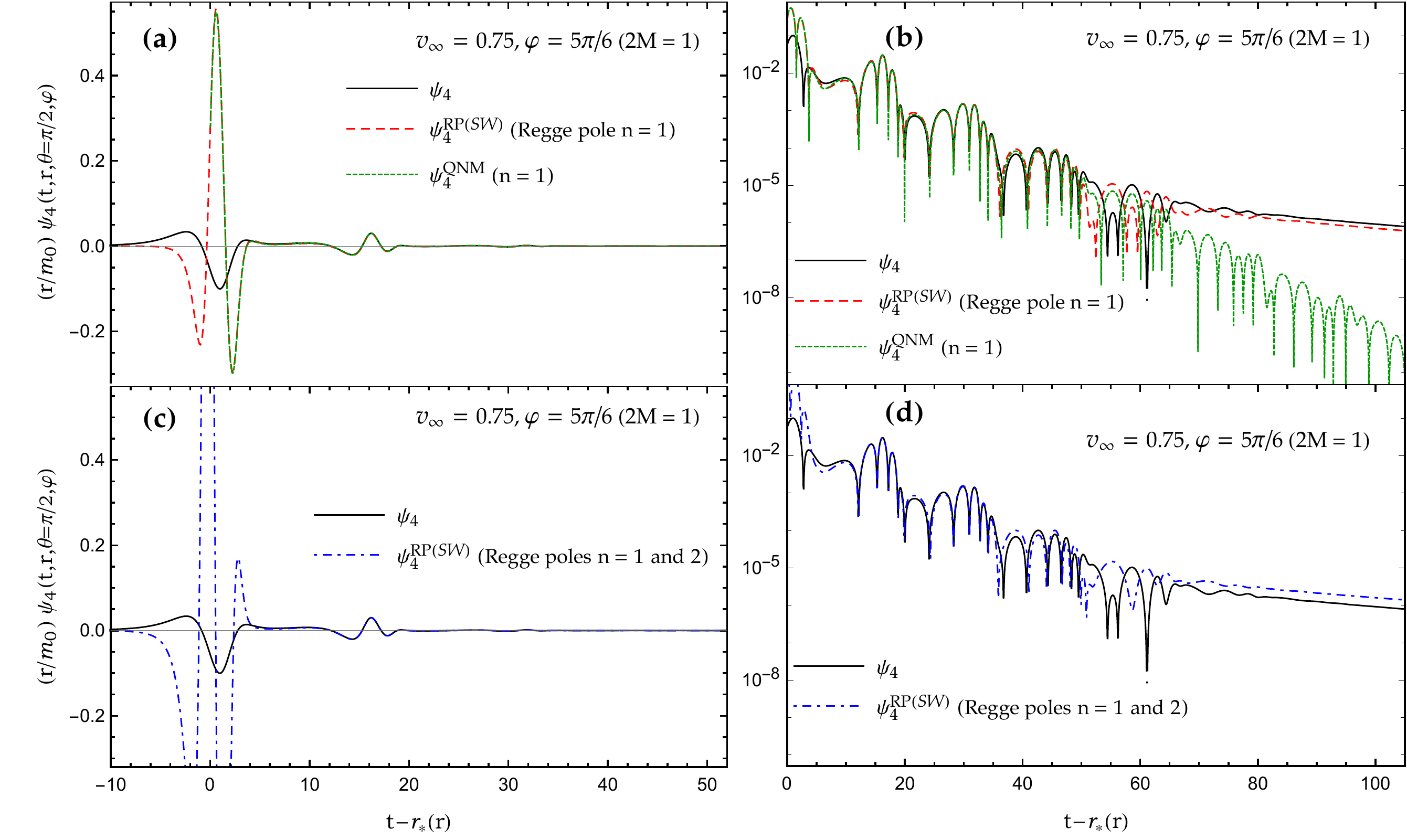}
\caption{\label{SW_Exact_QNM_CAM_5pis6_R} The Weyl scalar $\Psi_4$ and its Regge pole approximation $\Psi^{\text{\tiny{RP}} \, \textit{\tiny{(SW)}}}_4$ for $v_\infty =0.75$ ($\gamma \approx 1.51$) and $\varphi=5\pi/6$. (a) and (b) The pre-ringdown phase of the Weyl scalar $\Psi_4$ constructed by summing over the first thirteen partial waves is not described by the Regge pole approximation constructed from only one Regge pole. By contrast, this approximation matches very well the ringdown and the waveform tail. The quasinormal response $\Psi^\text{\tiny{QNM}}_4$ obtained by summing over the $(\ell, n)$ QNMs with $n=1$ and $\ell=2, \dots, 15$ is also displayed. At intermediate timescales, it matches very well the Regge pole approximation. (c) and (d) Taking into account an additional Regge pole does not improve the Regge pole approximation.}
\end{figure*}

 \begin{figure*}
\centering
 \includegraphics[scale=0.50]{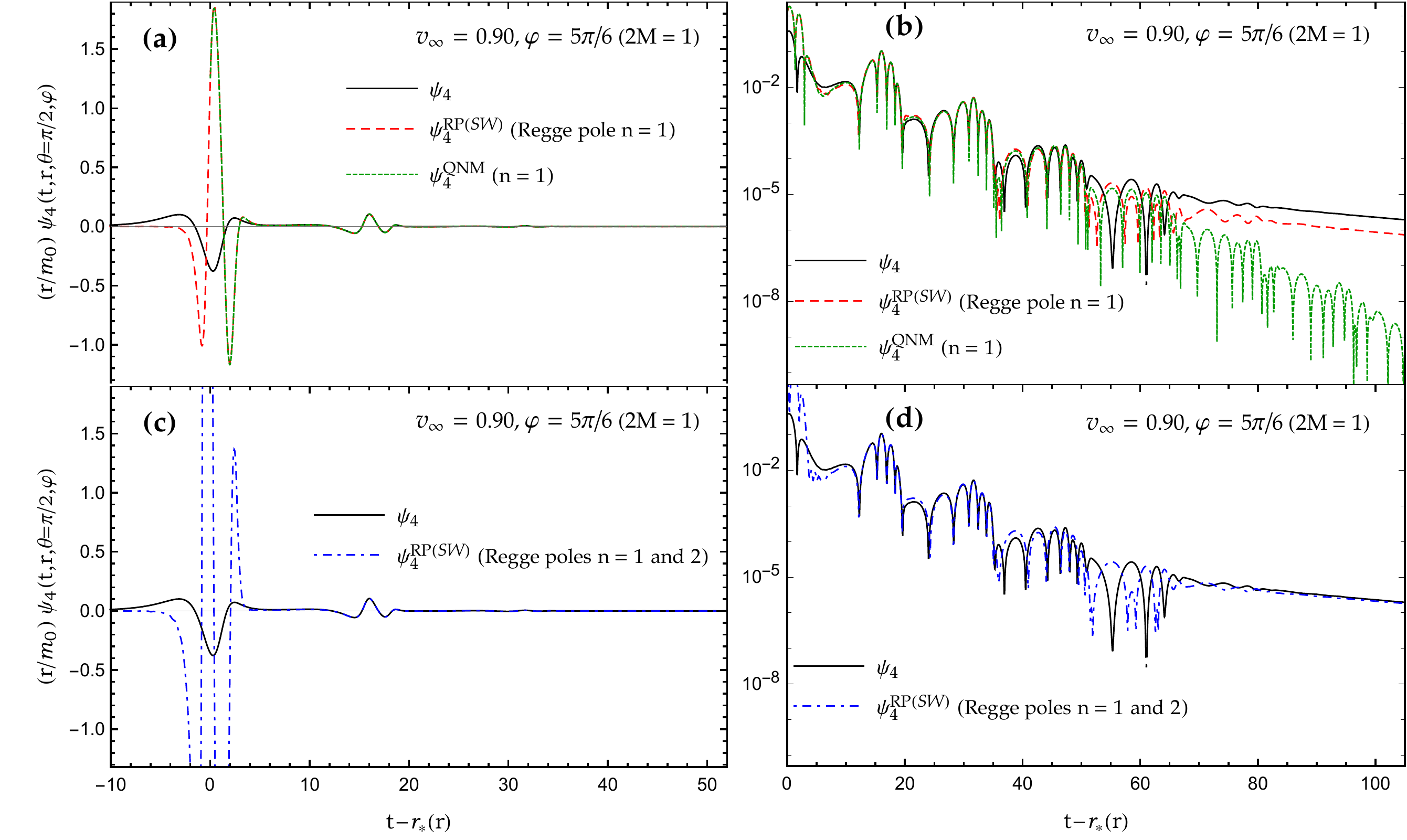}
\caption{\label{SW_Exact_QNM_CAM_5pis6_UR} The Weyl scalar $\Psi_4$ and its Regge pole approximation $\Psi^{\text{\tiny{RP}} \, \textit{\tiny{(SW)}}}_4$ for $v_\infty =0.90$ ($\gamma\approx 2.29$) and $\varphi=5\pi/6$. (a) and (b) The pre-ringdown phase of the Weyl scalar $\Psi_4$ constructed by summing over the first eighteen partial waves is not described by the Regge pole approximation constructed from only one Regge pole. By contrast, this approximation matches very well the ringdown and describes rather correctly the waveform tail. The quasinormal response $\Psi^\text{\tiny{QNM}}_4$ obtained by summing over the $(\ell, n)$ QNMs with $n=1$ and $\ell=2, \dots, 20$ is also displayed. At intermediate timescales, it matches very well the Regge pole approximation. (c) and (d) Taking into account an additional Regge pole only improves the description of the waveform tail.}
\end{figure*}

\subsection{Results and comments}
\label{SecIVb}

We have compared the multipolar waveform $\Psi_4$ given by (\ref{Psi4_ExpressionDef}) with its Regge pole approximation $\Psi^{\text{\tiny{RP}} \, \textit{\tiny{(P)}}}_4$ given by (\ref{CAM_Psi4_ExpressionDef_P_RP}) in Figs.~\ref{P_Exact_QNM_CAM_pis6}-\ref{P_Exact_QNM_CAM_pis2_R_avec_4} and with its Regge pole approximation $\Psi^{\text{\tiny{RP}} \, \textit{\tiny{(SW)}}}_4$ given by (\ref{CAM_Psi4_ExpressionDef_SW_RP}) in Figs.~\ref{SW_Exact_QNM_CAM_3pis4}-\ref{SW_Exact_QNM_CAM_5pis6_UR}. We have considered various values for the angle $\varphi \in [0, \pi]$ excluding the cases $\varphi=0$ and $\varphi=\pi$ for which $\Psi_4=0$. We have examined both the case of a particle starting at rest from infinity ($v_\infty =0$ and $\gamma=1$) and of a particle projected with a relativistic velocity at infinity [in that case, we have taken for the velocity at infinity $v_\infty =0.75$ ($\gamma\approx 1.51$) and $v_\infty =0.90$ ($\gamma\approx 2.29$)]. It is important to note that the number of partial modes to include in the sum (\ref{Psi4_ExpressionDef}) in order to obtain a numerically stable result strongly depends on the initial velocity of the particle: the sum over $\ell$ has been truncated at $\ell=10$ for $v_\infty =0$, at $\ell=15$ for $v_\infty =0.75$ and at $\ell=20$ for $v_\infty =0.90$.

As a preliminary remark, it should be recalled that, a long time ago, Davis, Ruffini and Tiomno \cite{Davis:1972ud} described the structure, as a function of the retarded time $t-r_\ast(r)$, of the partial waveforms $\psi_{\ell m}(t,r)$ [see our Eqs.~(\ref{TF_psi}) and (\ref{Partial_Response})] generated by the particle falling radially into a Schwarzschild BH. They identified three main components: a precursor (corresponding to the plunge of the particle from infinity), a burst (emitted when the particle approaches the BH photon sphere and crosses it) and a ringdown with a tail (emitted after the crossing of the photon sphere). In the present work, because we describe mathematically the BH response by the multipolar Weyl scalar $\Psi_4$ (i.e., by a superposition of a large number of partial waveforms), the part of the signal corresponding to the precursors and bursts of the various modes is blurred due to destructive interferences (see Figs.~\ref{P_Exact_QNM_CAM_pis6}-\ref{SW_Exact_QNM_CAM_5pis6_UR}). Hence, the two terms precursor and burst being now  inadequate, we will refer to the ``pre-ringdown phase'' to designate the early time response of the BH. By contrast, we can note that the multipolar Weyl scalar $\Psi_4$ still involves a ringdown at intermediate timescales and a tail at very late times.

In Figs.~\ref{P_Exact_QNM_CAM_pis6}-\ref{P_Exact_QNM_CAM_3pis4}, we have displayed the multipolar waveform $\Psi_4$ generated by a particle initially at rest at infinity and we have compared it with the Regge pole approximation $\Psi^{\text{\tiny{RP}} \, \textit{\tiny{(P)}}}_4$ obtained from the Poisson summation formula. In Figs.~\ref{P_Exact_QNM_CAM_pis6}-\ref{P_Exact_QNM_CAM_pis2}, for $\varphi=\pi/6, \pi/3$ and $\pi/2$ (let us note that these values of $\varphi$ are not too close to $\pi$ and recall that $\Psi^{\text{\tiny{RP}} \, \textit{\tiny{(P)}}}_4$ is divergent in the limit $\varphi \to \pi$), the approximation $\Psi^{\text{\tiny{RP}} \, \textit{\tiny{(P)}}}_4$ is in good or very good agreement with the waveform $\Psi_4$ even if we consider a single Regge pole and the agreement is even better if we consider an additional Regge pole. We can observe that the Regge pole approximation matches the pre-ringdown part of the signal as well as the ringdown and roughly describes the waveform tail. It is moreover important to note that it provides a description of the ringdown that does not necessitate determining a starting time. In Fig.~\ref{P_Exact_QNM_CAM_3pis4}, for $\varphi=3\pi/4$ (i.e., for a value of $\varphi$ rather close to $\pi$), the Regge pole approximation $\Psi^{\text{\tiny{RP}} \, \textit{\tiny{(P)}}}_4$ is no longer so interesting. Indeed, it only roughly describes the BH response. The result can be improved if an additional Regge pole is taken into account but we cannot be satisfied with the result. In fact, for values of $\varphi$ ``near'' $\pi$, it would be necessary to consider the background integral contribution $\Psi^{\text{\tiny{B}} \, \textit{\tiny{(P)}}}_4$ given by (\ref{CAM_Psi4_ExpressionDef_P_Background}) to correctly describe the multipolar waveform $\Psi_4$.

In Figs.~\ref{P_Exact_QNM_CAM_pis6_R}-\ref{P_Exact_QNM_CAM_pis2_R_avec_4}, for $\varphi=\pi/6, \pi/3$ and $\pi/2$, we have displayed the multipolar waveform $\Psi_4$ generated by a particle projected with a relativistic velocity at infinity and we have compared it with the Regge pole approximation $\Psi^{\text{\tiny{RP}} \, \textit{\tiny{(P)}}}_4$ obtained from the Poisson summation formula. We can observe that the Regge pole approximation is even more effective in the relativistic context. The whole signal is now impressively described. It should be noted that the relativistic particle excites many more QNMs than the particle initially at rest at infinity [it excites not only the fundamental $(\ell,n=1)$ QNMs but also their overtones] and that the Regge pole approximation permits us to capture their contribution efficiently.

In Fig.~\ref{SW_Exact_QNM_CAM_3pis4}, for $\varphi=3\pi/4$, we have displayed the multipolar waveform $\Psi_4$ generated by a particle initially at rest at infinity and we have compared it with the Regge pole approximation $\Psi^{\text{\tiny{RP}} \, \textit{\tiny{(SW)}}}_4$ obtained from the Sommerfeld-Watson transform. It should be recalled that, while $\Psi^{\text{\tiny{RP}} \, \textit{\tiny{(P)}}}_4$ constructed from the Poisson summation formula diverges in the limit $\varphi \to \pi$, this Regge pole approximation is regular in the same limit (it only diverges at $\varphi = 0$). As a consequence, it should provide better results than $\Psi^{\text{\tiny{RP}} \, \textit{\tiny{(P)}}}_4$ for $\varphi$ ``close to $\pi$''. This clearly appears if we compare Fig.~\ref{SW_Exact_QNM_CAM_3pis4} with Fig.~\ref{P_Exact_QNM_CAM_3pis4}. Now, the Regge pole approximation constructed from only one Regge pole does not describe the pre-ringdown phase of the Weyl scalar $\Psi_4$ but matches a large part of the ringdown and roughly approximates the waveform tail. In fact, in order to describe more correctly the multipolar waveform $\Psi_4$ and, in particular, the pre-ringdown phase, it might be necessary to consider the background integral contribution $\Psi^{\text{\tiny{B}} \, \textit{\tiny{(SW)}}}_4$ given by (\ref{CAM_Psi4_ExpressionDef_SW_Background}). It is moreover interesting to note that additional Regge poles do not improve the approximation. On the contrary, it seems that the Regge pole approximation $\Psi^{\text{\tiny{RP}} \, \textit{\tiny{(SW)}}}_4$, as a series over the Regge poles, diverges.

In Figs.~\ref{SW_Exact_QNM_CAM_5pis6}-\ref{SW_Exact_QNM_CAM_5pis6_UR}, for $\varphi=5\pi/6$ , we have displayed the multipolar waveform $\Psi_4$ generated by a particle initially at rest at infinity and by a particle projected with a relativistic velocity and we have compared it with the Regge pole approximation $\Psi^{\text{\tiny{RP}} \, \textit{\tiny{(SW)}}}_4$ obtained from the Sommerfeld-Watson transform. Here again, the Regge pole approximation constructed from only one Regge pole does not describe the pre-ringdown phase of the Weyl scalar $\Psi_4$ but it matches a large part of the ringdown and approximates rather correctly the waveform tail. The Regge pole approximation is even more effective in the relativistic context in which the ringdown and the tail are now impressively described. It should be noted that it is necessary to limit the number of Regge poles involved in the sum defining $\Psi^{\text{\tiny{RP}} \, \textit{\tiny{(SW)}}}_4$ due to the divergence of this series.

\section{Conclusion}
\label{Conc}

In this article, by using CAM techniques, we have revisited the problem of the gravitational radiation generated by a massive particle falling radially from infinity into a Schwarzschild BH. More precisely, we have described the emitted gravitational waves by the Weyl scalar $\Psi_4$ and we have extracted from its multipole expansion (\ref{Psi4_ExpressionDef}), as an approximation, the Fourier transform of a sum over the Regge poles of ${\cal S}$-matrix of the BH also involving their residues (or, more exactly, the excitation factors of the Regge modes). In fact, we have obtained two different Regge pole approximations: the first one, which is given by (\ref{CAM_Psi4_ExpressionDef_P_RP}), has been constructed from the Poisson summation formula and provides very good results for observation directions in a large angular sector around the particle trajectory; the second one, which is given by (\ref{CAM_Psi4_ExpressionDef_SW_RP}), has been constructed from the Sommerfeld-Watson transform and provides good results in a large angular sector around the direction opposite to the particle trajectory. By using these two Regge pole approximations, we have clearly highlighted the benefits of working within the CAM framework, and we now briefly summarize the main results we have numerically obtained concerning the structure of the Weyl scalar $\Psi_4$:
\begin{enumerate}[label=(\arabic*)]

\item In general, the Regge pole approximation constructed from only one Regge pole describes very well the ringdown or a large part of the ringdown. In certain cases, the agreement is impressive. Contrary to the QNM description of the ringdown, the Regge pole description does not necessitate a starting time. Moreover, if a very large number of QMNs are excited (this is the case if the particle is projected with a relativistic velocity), the Regge pole description can be improved by considering a few additional poles, which permits us to fit the BH response taking into account the ``quasinormal overtones''.

\item In general, the Regge pole approximation roughly describes the waveform tail and, in certain circumstances (this is the case if the particle is projected with a relativistic velocity), the description is very good.

\item For observation directions close to the particle trajectory, the Regge pole approximation can be used to describe the pre-ringdown part of the BH response, i.e., it matches the superposition of all the precursors and bursts involved in the partial wave expansion.

\end{enumerate}
We have, moreover, noted that the multipolar waveform $\Psi_4$ can be entirely described if we add to the Regge pole approximation (\ref{CAM_Psi4_ExpressionDef_P_RP}) the background integral contribution (\ref{CAM_Psi4_ExpressionDef_P_Background}) or to the Regge pole approximation (\ref{CAM_Psi4_ExpressionDef_SW_RP}) the background integral contribution (\ref{CAM_Psi4_ExpressionDef_SW_Background}). In this article, we have not taken into account these background integral contributions. They could be helpful to improve the CAM description of the gravitational radiation generated by the particle. For this purpose, it would be interesting to evaluate them numerically or asymptotically and to provide a physical interpretation of the results.

It is important to recall the philosophy underlying the use of CAM techniques in the context of BH physics (see also the Introduction). The information lying in the ${\cal S}$-matrix of the BH can be extracted in two alternative ways: by considering the analytic structure of the ${\cal S}$-matrix in the complex $\omega$ plane or in the CAM plane. So, we can consider that the information encoded in the QNM spectrum (taking, in addition, into account the branch cut of the ${\cal S}$-matrix elements) and that encoded in the Regge-mode spectrum are equivalent. As a consequence, it is natural to be able to describe, from a Regge pole analysis, the ringdown and the waveform tail of the emitted gravitational radiation. Moreover, it is worth noting that the high-frequency behavior of the Regge poles has permitted us to interpret semiclassically the weakly damped QNMs \cite{Decanini:2002ha,Decanini:2009mu,Decanini:2010fz,Decanini:2011xi}. In this article, we have increased the role of the high-frequency contributions by describing the gravitational radiation from the Weyl scalar $\Psi_4$ (i.e., from a field) instead of using the metric perturbations $h_+$ and $h_\times$ (i.e., a potential) with, as a consequence, interesting results provided by the Regge pole approach.

The alternative description of gravitational radiation from BHs we have proposed in this article could play a fundamental role in gravitational-wave physics. But, of course, in order to establish this definitively, it would be interesting to go beyond the relatively simple problem examined here (as well as to address unresolved issues). In particular, it would be interesting to consider the gravitational radiation generated by a particle with an arbitrary orbital angular momentum plunging into a Schwarzschild or a Kerr BH.

\begin{acknowledgments}

We wish to thank Yves Decanini and Julien Queva
for various discussions and Paul Gabrielli for providing us with powerful computing resources.

\end{acknowledgments}

%

\bibliography{RP_RT_GW}

\end{document}